\documentclass{ws-procs9x6}

\begin{document}

\title{SPIN GLASS\\
A BRIDGE BETWEEN QUANTUM COMPUTATION AND STATISTICAL MECHANICS}
\author{MASAYUKI OHZEKI}

\address{Department of Systems Science, Graduate School of Informatics, Kyoto University,\\
Yoshida-Honmachi, Sakyo-ku, Kyoto 606-8501, Japan\\
E-mail: mohzeki@i.kyoto-u.ac.jp\\
http://www-adsys.sys.i.kyoto-u.ac.jp/mohzeki/}

\begin{abstract}
In this chapter, we show two fascinating topics lying between quantum
information processing and statistical mechanics. First, we introduce an
elaborated technique, the surface code, to prepare the particular quantum
state with robustness against decoherence. Interestingly, the theoretical
limitation of the surface code, accuracy threshold, to restore the quantum
state has a close connection with the problem on the phase transition in a
special model known as spin glasses, which is one of the most active
researches in statistical mechanics. The phase transition in spin glasses is
an intractable problem, since we must strive many-body system with
complicated interactions with change of their signs depending on the
distance between spins. Fortunately, recent progress in spin-glass theory
enables us to predict the precise location of the critical point, at which
the phase transition occurs. It means that statistical mechanics is
available for revealing one of the most interesting parts in quantum
information processing. We show how to import the special tool in
statistical mechanics into the problem on the accuracy threshold in quantum
computation.

Second, we show another interesting technique to employ quantum nature,
quantum annealing. The purpose of quantum annealing is to search for the
most favored solution of a multivariable function, namely optimization
problem. The most typical instance is the traveling salesman problem to find
the minimum tour while visiting all the cities. In quantum annealing, we
introduce quantum fluctuation to drive a particular system with the
artificial Hamiltonian, in which the ground state represents the optimal
solution of the specific problem we desire to solve. Induction of the
quantum fluctuation gives rise to the quantum tunneling effect, which allows
nontrivial hopping from state to state. We then sketch a strategy to control
the quantum fluctuation efficiently reaching the ground state. Such a
generic framework is called quantum annealing. The most typical instance is
quantum adiabatic computation based on the adiabatic theorem. The quantum
adiabatic computation as discussed in the other chapter, unfortunately, has
a crucial bottleneck for a part of the optimization problems. We here
introduce several recent trials to overcome such a weakpoint by use of
developments in statistical mechanics.

Through both of the topics, we would shed light on the birth of the
interdisciplinary field between quantum mechanics and statistical mechanics.
\end{abstract}


\bodymatter

\section{Introduction: Statistical Mechanics and Quantum Mechanics}

\label{aba:sec1} Quantum mechanics is a method partially based on the
concept of the probability in outputs of the measurements on the physical
state. This is because the physical state is allowed to be expressed by
superposition of the possible eigenstates with the probability amplitude,
which can be determined by the Schr\"odinger equation. Readers might feel
uneasy since such a probabilistic phenomena seems to be problematic and
uncontrollable. However, as shown in this chapter, several techniques have
been designed and proposed for realization of the stability in quantum
systems.

The quantum state is fragile, sensitive to noisy environment effects and
continues to change. This peculiar destruction of the quantum state is known
as decoherence. If we control to maintain the original state, we have to
cure the quantum state suffering from undesired errors due to decoherence.
The technique to remove these errors is the quantum error correction, which
occupies the first part of this chapter. To fix successfully the quantum
state, we need to prepare the redundant degrees of freedom to restore the
original property and check where and how errors exist. We must thus design
and deal with a many-body quantum system, which is known as an intractable
problem. However it is worthwhile to strive such a difficult problem. Once
you obtain an ingenious way to keep the quantum state, you find a pavement
toward the realization of the quantum information processing, since you can
encode simultaneously multiple information in the single quantum state by
use of superposition. We here introduce a quantum error correcting
technique, surface code \cite{Kitaev}, which skillfully use the concept of
the topology to encode the original state in quantum many-body system. The
surface code has a remarkable feature closely related with the main concept
of this chapter. The accuracy threshold, which represents the theoretical
limitation to restore the original quantum state by the surface code, is
related with the phase transition in the special model in spin glasses \cite%
{Dennis}. Spin glass is a disordered magnetic material, which shows a
peculiar behavior with extraordinary slow relaxation toward equilibrium
state in a low temperature \cite{Rev1,Rev2,Rev3,MPV,HNbook}. The complicated
interactions in spin glasses spoil several methods used for systematic
analyses in statistical mechanics. The identification of the critical point
in spin glasses is used to be an intractable problem for long days.
Fortunately, the recent development of theory in spin glasses yields a
systematic analysis for its precise location. It means that the specialized
tool to analyses in spin glasses is available for the problem in quantum
computation. We show the fascinating connection between two unrelated
topics, while reviewing the systematic approach to derive the accuracy
threshold.

By using such a technique to maintain the quantum state as introduced above,
we can prepare a coherent state. Then, how do we use superposition of the
quantum system? Superposition enables us to evolve the quantum system in a
parallel way over the various possible states. A famous algorithm given by
Shor \cite{QC1} is also based on the simultaneous search over all the
possible candidates. In the second part of the present chapter, we introduce
a generic technique by use of quantum nature in this chapter, called quantum
annealing \cite{QA1,QA2,QA3,QA4,QA5,QA6,QA,QAMN}. Quantum annealing is an
active use of superposition in order to obtain the most important result by
searching all the candidates. Its attractive feature is simplicity and
generality in applications. This is useful to solve several particular
issues known as the optimization problems, in which we desire to find a
minimizer or maximizer of the given multivariable function \cite{OP1,OP2}.
The Shor's algorithm uses a skillful technique in order to efficiently
provide a desired answer. On the other hand, quantum annealing basically
takes a simpler way only by controlling quantum fluctuations. The most
typical procedure of quantum annealing, quantum adiabatic computation, is
superior to its classical counterpart, simulated annealing \cite{SA1,SA2},
owing to usage of quantum fluctuation. However a bottleneck of quantum
adiabatic computation is also revealed in the application to the particular
optimization problems. This weak point can be understood thorough a special
mapping from the classical stochastic time evolution to the particular
quantum dynamics. We are again between statistical mechanics and quantum
information processing. Therefore, in this chapter, we will go further to
overcome the bottleneck of quantum adiabatic computation by use of this
fascinating relationship. We show several attempts for surmounting the
obstacle involved in quantum adiabatic computation by employing several
alternative strategies from statistical mechanics.

Statistical mechanics uses the probability similarly to quantum mechanics,
but is capable to predict a definite behavior in the future in a
large-number of components called as thermodynamic limit. The mathematical
background of statistical mechanics, large deviation, provides the ability
to give a definite answer even by use of the probability. Observant readers
can find out the key point in this chapter. That is to deal with a large
number of components. Turning on our eyes on substances around our daily
life, they shows stable appearance. Material consists of many components,
atoms, molecules and their mixtures, which should follow the rules of the
quantum mechanics. The stability of these macroscopic systems comes from the
particular property of the collection of many components. However, in many
body systems, once if you tune the external parameters as temperature,
pressure, and some fields, they can eventually change their uniform
appearances. For instance, increase of temperature makes change of solid
into liquid and gas. This phenomena is known as the phase transition. The
phase transition involves singularities in the behavior of the physical
quantities. The both of quantum error correction and quantum annealing shown
in this chapter, suffer from this peculiar behavior. The failure to keep the
quantum state by the surface code and the decay of the performance in
quantum adiabatic computation come from the properties of the phase
transition. Therefore, by dealing with the problems on the phase transition,
we study the fascinating interdisciplinary connection between statical
mechanics and quantum information processing.

Recent developments in manufacturing and manipulation of the quantum
mechanical system enables us to prepare a large number of components. In
this sense, the topics we deal with in this chapter must be valuable for
developments in actual applications in the future.

\section{Training: Statistical Mechanics}

For unfamiliar readers with statistical mechanics and many students to
understand the essential parts in this chapter, let us get back to the
starting point of statistical mechanics.

At first let us recall the concept of the probability. What was the
probability? If you have a non-tricky coin, you can say that the probability
you can find a face (head or tail) should be a half. What does it mean?

\subsection{Student's misreading point: probability is...}

We use an artificial variable with a binary $S=\pm 1$ to represent the coin
state as head ($+1$) or tail ($-1$). A non-tricky coin takes the probability
as $P_1(S=1)=1/2$ and $P_1(S=-1)=1/2$. We also can deal with a tricky coin
with a biased probability as 
\begin{equation}
P_1(S) = \frac{\exp(K S)}{Z_1(\beta)},  \label{1Prob}
\end{equation}
where $K$ is the strength of bias, and $Z_1(K)$ is a normalized constant
explicitly given as $2\cosh K$. By use of the above probability, we can
evaluate two characteristic quantities of the probabilistic system. The
first quantity is expectation expressed as $m$, which indicates tendency of
the coin whether head or tail, 
\begin{equation}
m = \sum_{S=\pm 1}S P_1(S) = \tanh(K).
\end{equation}
It readily shows that the coin state tends to be head if we increase the
bias of the tricky coin $K \to \infty$. Another is the variance defined
through the square of the difference from the expectation as 
\begin{equation}
\sigma^2 = \sum_{S}(S-m)^2 P_1(S) = 1-\tanh^2(K).
\end{equation}
If we increase the bias $K$, the variance can vanish. However this is the
case you know that this coin must be head without any probabilistic factors
because $P_1(S=-1) \to 0$.

Then, if you once glance at the result of the single trial, can you judge
whether its behavior is closely related with the obtained value of $m$ or
not? The answer is "No". The value of $m$ only gives "tendency". The actual
results will fluctuate around the expectation in repeating observations. We
must repeat the coin games many times.

The above is the usual statement around the explanation of the probability.
Since students, who studied the probability theory or other related topics,
do not find out its genuine meaning, they often say that ``probability" is
obscure, abstract, and difficult!!, and conclude that it is not a
well-understandable concept. Unfortunately, quantum mechanics and
statistical mechanics with the concept of probability might be considered as
non-reliable and non-deterministic to predict the future-coming behavior of
the system we deal with. I would like to ask you why the probability for a
non-tricky coin was a half. The students must recognize what the expectation
express in our future. As you know, the expectation is not relevant for
understanding the output in a single coin game. Let us play the game
repeatedly. Then...

\subsection{Probability describes... a certain behavior}

In order to clarify the concept of the probability, in particular
expectation, let us consider to accumulate the results of $N$-time
observations after flipping coins. In all the trials, their results are
characterized by the individually and independent distribution, which is
same as in Eq. (\ref{1Prob}). Gradually approaching the specific example of
use of probability in statistical mechanics, we take a simple model of the
magnetic material, which is the typical example of studies in this field.

Let us introduce the Ising variable, which describes the magnetic momentum
of spins in the magnetic material, often simply termed as ``spin" or ``Ising
spin". The Ising spin takes only two integers as $\pm 1$. We here regard the
behavior of $N$-independent spins as the flipping coins in the $N$%
-sequential times. In statistical mechanics, we choose the specific
probability distribution depending on the conditions of the system under
consideration in order to calculate the expectation. Notice that, in
statistical mechanics, the probability distribution can be given "\textit{a
priori}" similarly to the above case of a non-tricky coin. In the previous
simple case, we could choose a half as the probability for the coin since it
was reasonable. Generally we do not know the precise structure of the
probability distribution for various events. On the other hand, statistical
mechanics provides us with the probability distribution of the behavior in
equilibrium state. For instance, for the equilibrium state without change of
the number of components, we use the canonical distribution characterized by
the Hamiltonian and temperature. 
\begin{equation}
P(S_1,S_2,\cdots,S_N) = \frac{1}{Z_N(\beta)}\exp\left(-\beta
H(S_1,S_2,\cdots,S_N)\right),
\end{equation}
where we define the inverse temperature $\beta=1/k_BT$ (often $k_B=1$) and $%
Z_N(\beta)$ is called as the partition function. The Hamiltonian describes
the energy depending on the microscopic state of the system. The temperature
controls thermal fluctuation to drive microscopic degrees of freedom in the
system. If $\beta \to 0$, the spins are not stable due to strong effect by
thermal fluctuation. On the other hand, in the low temperature $\beta \to
\infty$, the system is settled into a lower energy state. The single spin
favors parallel direction to the magnetic field since its energy can be
described by the Hamiltonian $H(S) = - h S$, where $h$ is the strength of
the magnetic field. Then the joint probability of the $N$ spins ($N$-time
coin games) can be written by the canonical distribution as 
\begin{equation}
P_N(S_1,S_2,\cdots,S_N) = \frac{1}{Z_N(\beta)}\prod_{i=1}^{N} \exp(\beta h
S_i),  \label{Joint_Prob}
\end{equation}
where 
\begin{equation}
Z_N(\beta) = \left(2\cosh(\beta h )\right)^N.
\end{equation}
The partition function plays a roll of the characteristic function. Its
logarithmic function is called the free energy $-\beta F_N(\beta) = \log
Z_N(\beta)$. We often use the thermal average by use of the canonical
distribution in statistical mechanics written as 
\begin{equation}
\langle O \rangle = \sum_{\{S_i\}} O(\{S_i\}) P_N(S_ ,S_2,\cdots,S_N).
\end{equation}
where $O$ denotes the physical observable.

Let me ask the following question. If you have huge number of the outputs in
an experiment, how do you interpret the result. Usually we use the average
over all the realizations to simplify the outputs. This simplified picture
is often called as coarse graining in statistical mechanics. What we must
take care of is that the average (sample mean) over all the realizations 
\begin{equation}
m_{N} = \frac{1}{N} \sum_{i=1}^{N} S_i,
\end{equation}
and its variance 
\begin{equation}
\sigma_{N} = \frac{1}{N} \sum_{i=1}^{N} (S_i- m_N)^2.
\end{equation}
Statistical mechanics gives a definite future of the above average for the
many-body system in terms of the expectation and variance estimated from the
specific distribution function.

Below let us see the connection between quantities given by the empirical
measure and those by the \textit{a priori} distribution function.

\subsection{Large deviation property}

\begin{figure}[tb]
\begin{center}
\includegraphics[angle=-90,width=100mm]{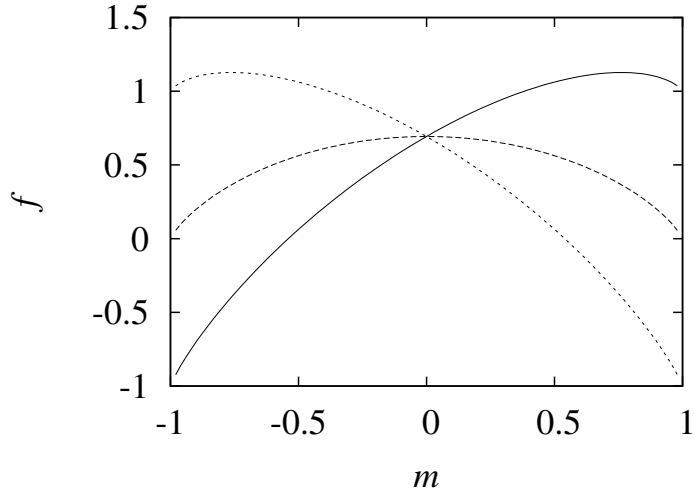}
\end{center}
\caption{{\protect\small Behavior of $f(m,\tilde{m}^*)$ with $\protect\beta %
h = 1.0$ (solid), $0$ (thick symmetric dashed curve), and $-1.0$ (dotted
curve).}}
\label{Prob_m}
\end{figure}

We do not care the detail on the system with the large-number components.
For instance, for the above $N$ spin systems, it is enough to characterize
the magnetic property by the average of the $N$ Ising spins. This is called
as the magnetization, which indicates the magnetic strength of the material.
Statistical mechanics predicts such significant quantities in the
macroscopic scale by averages over all the components, coarse-grained
quantities. If we consider the large-number limit $N \to \infty$, the
averaged values can correspond to the expectations estimated by the
probability distribution under a certain condition shown below. It is
convenient to change the expression of the joint probability (\ref%
{Joint_Prob}) into the form of the probability of the average as 
\begin{equation}
P_N(m) = \sum_{\{S_i\}} \delta\left(m - \frac{1}{N}\sum_{i=1}^NS_i%
\right)P_N(S_1,S_2,\cdots,S_N).
\end{equation}
By use of the integral form of the delta function as 
\begin{equation}
\delta\left(x\right) = \int^{\mathrm{i} \infty}_{-\mathrm{i} \infty} d\tilde{%
x} \exp(x\tilde{x}),
\end{equation}
we can obtain 
\begin{equation}
P_N(m) \propto \sum_{\{S_i\}} \int d\tilde{m}\exp \left\{ N m\tilde{m} +
\left(-\tilde{m}\sum_{i=1}^NS_i + \beta h \sum^N_{i=1}S_i \right) \right\}.
\end{equation}
Before performing the integration, we sum over the spin variables. 
\begin{equation}
P_N(m) \propto \int d\tilde{m}\exp \left\{ N m\tilde{m} + N \log \left( 2
\cosh \left(\beta h - \tilde{m}\right)\right) \right\}.
\end{equation}
The integrand is proportional to $N$. In that case, we can apply the
saddle-point method to the above integration. When we consider the infinite
limit of $N$, the integral is given by the maximizer $\tilde{m}^*$ of the
integrand as 
\begin{equation}
P_N(m) = \exp \left\{ -N f(m,\tilde{m}^*) \right\}.
\end{equation}
Here we define the pseudo free energy $f(m,\tilde{m})$ as 
\begin{equation}
f(m,\tilde{m}) = - m\tilde{m} - \log \left( 2 \cosh \left( \beta h - \tilde{m%
}\right)\right).
\end{equation}
The saddle-point equation for $\tilde{m}$, which gives the maximum of the
integrand (or the minimum of the pseudo free energy), is 
\begin{equation}
\frac{\partial f}{\partial \tilde{m}} = 0 \to m = \tanh \left(\beta h-\tilde{%
m}\right) .
\end{equation}
The maximizer of the integrand is given by $\tilde{m}^*= \beta h -
\tanh^{-1}(m)$. By use of this maximizer, we obtain the probability of the
magnetization as 
\begin{equation}
P_N(m) = \exp \left\{ N \left( m \beta h - m \tanh^{-1} m + \log \frac{2}{%
\sqrt{1-m^2}}\right) \right\}.
\end{equation}
In Fig. \ref{Prob_m}, we describe the behavior of $f(m,\tilde{m}^*)$. 
The maximum is located at $m^* = \tanh(\beta h)$, which can be verified by
derivative with respect to $m$. If $N$ increase, the probability away from $%
m^* = \tanh(\beta h)$ decrease exponentially. The maximum indicates the
value of the expectation since 
\begin{equation}
\langle m \rangle = \left\langle \frac{1}{N} \sum_{i=1}^N S_i \right\rangle
= \frac{1}{N}\frac{\partial P_N(m)}{ \partial \hat{m}} = \tanh(\beta h).
\end{equation}
Moreover, if we evaluate the variance, we reach 
\begin{equation}
\langle \sigma^2_N \rangle = \left\langle \frac{1}{N} \sum_{i=1}^N ( S_i
\right\rangle =\frac{1}{N}\frac{\partial^2 P_N(m)}{\partial \hat{m}^2} = 
\frac{1}{N}\left(1-\tanh^2 (\beta h)\right).
\end{equation}
Therefore the magnetization can take a ``definite" value $m^* = \tanh(\beta
h)$ if $N \to \infty$. This property is known as the large deviation
principle in mathematics. In the above simple case, we treat the
individually-independent distribution. Fortunately, it is confirmed that the
large deviation principle holds in relatively wide varieties rather than
individually-independent distribution. Statistical mechanics plentifully
uses explicitly and implicitly this fruitful property. It means that, in
many-body system, we can predict a definite future even by use of the
probability as classical mechanics with the deterministic equation. In other
words, statistical mechanics predicts the average value for the physical
observables of the system with large number components from the calculation
of the expectation by use of an \textit{a priori} distribution function such
that the canonical, micro canonical, and ground canonical ensemble.

\subsection{Mean-field analysis}

The probability can give a definite future expressed by the expectation
value, if we are interested in many-body systems by the support of the large
deviation principle. However many degrees of freedom in material strongly
interact with each other. The spin is not an exception. Differently from the
above simplified case, we must take care of the effect through
magnetic-dipole interactions to more precisely understand the behavior of
magnetic material. We exemplify its simplest model to deal with effects of
interactions known as the Ising model, whose Hamiltonian is 
\begin{equation}
H = - J \sum_{\langle ij \rangle} S_i S_j.  \label{IsingH}
\end{equation}
The subscript stands for the index of the site, where the spin is located.
The summation is taken over the nearest-neighboring pairs of spins. The
geometric property as the locations of the spins is closely related with the
structure of the magnetic material. As precisely as possible we analyze an
actual behavior of magnetic material, we must consider a finite-dimensional
structure of atoms. However it is difficult to straightforwardly perform
nontrivial analyses on the finite-dimensional Ising model. We take several
approximations or numerical simulations for obtaining meaningful results of
the finite-dimensional Ising models. The exactly solvable examples without
any approximations are found in one and two dimensions but rare exceptions.

\begin{figure}[tb]
\begin{center}
\includegraphics[angle=-90,width=100mm]{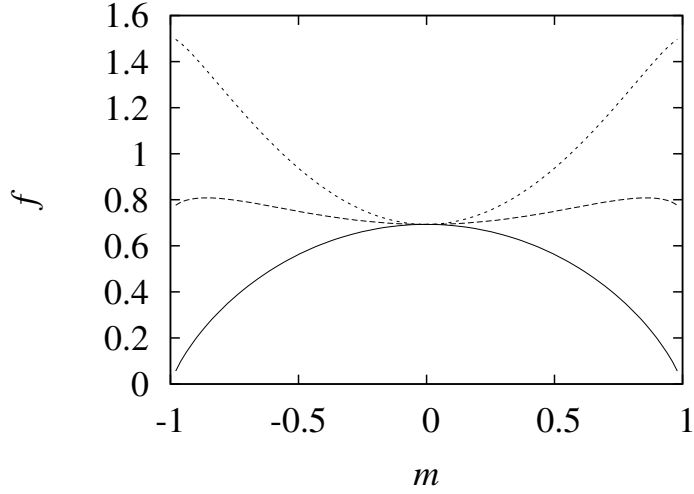}
\end{center}
\caption{{\protect\small Behavior of $f(m,\tilde{m}^*)$ for $z=6$, while we
set $J=1$, with $\protect\beta = 0.0$ (solid), $0.1$ (dashed curve), and $%
0.25$ (dotted curve).}}
\label{Prob_mMF}
\end{figure}

\begin{figure}[tb]
\begin{center}
\includegraphics[angle=-90,width=100mm]{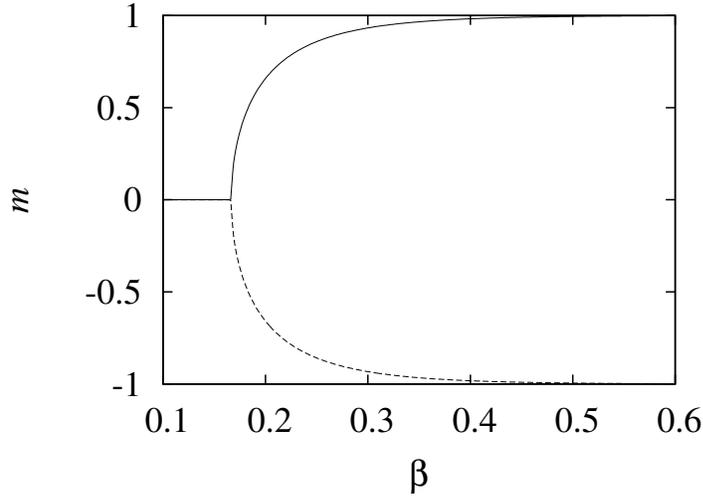}
\end{center}
\caption{{\protect\small Magnetization given by the mean-field analysis for
the Ising model $z=6$. Here we set $J=1$.}}
\label{MF}
\end{figure}

The simple but qualitative approximation often taken as the first trial is
the mean-field analysis. It enables us to obtain a meaningful picture to
describe a peculiar behavior of the many-body system. The probability
distribution of the degrees of freedom should be correlated with each other
in the case of the existence of the interactions. However we simply assume
that the probability distribution should be factorized as
individually-independent one as in Eq. (\ref{Joint_Prob}). As seen above,
under this approximation, we can characterize the system by use of the
magnetization $m$. In other words, the surrounding spins adjacent to a
particular spin being replaced by a uniform variable $m$, we convert the
many-body problem into a simple one-body system. Let us rewrite the
Hamiltonian by the replacement of the surrounding spins $S_j = m$ to $S_i$
as 
\begin{equation}
H_{\mathrm{MF}} (S_i)= - \frac{zJ m}{2}\sum_{i=1}^N S_i,
\end{equation}
where $z$ is the coordination number (usually in the case of the hyper cubic
system, $z=2d$, where $d$ is the dimension). Similar calculations to the
previous example give rise to the final expression of the probability for $m$
as 
\begin{equation}
P_N(m) \propto \int d\tilde{m}\exp \left\{ N m\tilde{m} + N \log \left( 2
\cosh \left(\frac{\beta Jz}{2} m- \tilde{m}\right)\right) \right\}.
\end{equation}
The saddle point equation is 
\begin{eqnarray}
\frac{\partial f}{\partial \tilde{m}} = 0 & \to & m = \tanh \left(\beta 
\frac{Jz m}{2}-\tilde{m}\right) .
\end{eqnarray}
Thus we find the probability of the magnetization for the interacting system 
\begin{equation}
P_N(m) \propto \exp \left\{ N \left(\frac{\beta J z}{2} m^2 - m \tanh^{-1} m
- \log \frac{2}{\sqrt{1-m^2}}\right) \right\}.  \label{m_Prob}
\end{equation}
We describe the behavior of the free energy as in Fig. \ref{Prob_mMF} 
Readers can find two maximizers in the low-temperature region. Let us
evaluate explicitly this unexpected result. The most probable state is given
by the maximizer of the following self-consistent equation 
\begin{equation}
m = \tanh (\beta Jz m).
\end{equation}
The explicit value of the magnetization over all temperature is shown in
Fig. \ref{MF}. 
Beyond the special point $\beta = 1/zJ$, the magnetization suddenly take
non-zero value despite of absence of magnetic field. This is the spontaneous
symmetry breaking, in other terms, the phase transition. This is also a
peculiar property in many-body systems. The mean-field analysis gives
several meaningful properties including the phase transition on many-body
systems as above. The results by the mean-field analysis can be validated in
infinite dimensions and reflect on the behavior in such higher dimensions.
Therefore we must take care of their applicability to the actual behavior of
the many-body system in finite dimensions.

\subsection{Phase transition}

Let us more qualitatively discuss the phase transition through the results
obtained by mean-field analysis. Without the magnetic field, the spins do
not have any tendency of the direction, that is $m \approx 0$. Indeed the
most probable state is given by $m=0$. This solution describes the
paramagnetic phase of the magnetic material without any magnetization.
However if you expand the self-consistent equation under the assumption that 
$m$ should be small as 
\begin{equation}
m = \beta Jz m.
\end{equation}
We find the special temperature $\beta =1/zJ$, at which the magnetization $m$
can take a non-zero value. This is a signature of the phase transition from
the paramagnetic state into the ferromagnetic state. We expand the
self-consistent equation up to third order of $m$, and thus obtain 
\begin{equation}
m = \beta Jz \left\{ m - \frac{1}{3} \left( \beta Jz \right)^2 m^3\right\}.
\end{equation}
This equation indeed gives the ferromagnetic solutions with non-zero
magnetization as $m = \pm\sqrt{3(1-\beta J z)}/(\beta J z)$, which vanish at 
$\beta =1/zJ$. This special temperature is called the critical point. In
addition, two phases, paramagnetic and ferromagnetic phases are
characterized by the behavior of the magnetization. Such a quantity, which
can distinguish the phase, is called as the order parameter. Could you find
the fact that you have two ferromagnetic solutions of the magnetization? It
implies that the two maximizers, say two probable states, are allowed to
appear in Eq. (\ref{m_Prob}). Then, which solution should be chosen and will
appear naturally in experience? It depends on the dynamics rule and the
initial conditions. If you design the dynamics to simulate thermal
fluctuations At least, we can find the fact that, in the ferromagnetic
phase, state with $m=0$ would not be expected to occur in the case with a
large $N$. That means that two of the probable states are separate and can
not change one into another in the actual experience. In other words, the
system has some robustness against thermal fluctuations. That will be an
analogous property to protect the fragile quantum state later.

\subsection{Spin Glasses}

The above exemplified model have a uniform interactions between the Ising
spins. The theoretical model of spin glasses is usually not the case. The
simplest model, the Edwards-Anderson model, is defined by the following
Hamiltonian 
\begin{equation}
H = - \sum_{\langle ij \rangle} J_{ij}S_iS_j,
\end{equation}
where $J_{ij}$ is the disordered interactions assumed to obey several type
of the distribution functions. We often deal with two typical cases, which
take a bimodal distribution (then called as $\pm J$ Ising model) 
\begin{equation}
P(J_{ij}) = p \delta(J_{ij} - J) + (1-p)\delta(J + J_{ij}),  \label{pmJ}
\end{equation}
and the Gaussian distribution with the average $J_0$ and the variance $J$ 
\begin{equation}
P(J_{ij}) = \frac{1}{\sqrt{2\pi J^2}}\exp\left(- \frac{1}{2J^2}(J_{ij} -
J_0)^2 \right).
\end{equation}
The partition function should be dependent of the specific configuration as 
\begin{equation}
Z(\beta;\{J_{ij}\}) = \sum_{\{S_i\}} \prod_{\langle ij \rangle} \exp \left(
\beta J_{ij} S_iS_j\right).
\end{equation}
In order to evaluate the partition function and thus the free energy, we
strive the difficult task to deal with the non-uniform interactions.
Instead, we usually take a wise strategy to evaluate the averaged free
energy based on the self-averaging property as, in the large-limit $N$, 
\begin{equation}
\frac{1}{N} F(\beta;\{J_{ij}\}) \to \frac{1}{N} \left[ F(\beta;\{J_{ij}\}) %
\right],
\end{equation}
where the square bracket denotes the average over all the combinations of $%
\{J_{ij}\}$ (configurational average) and the free energy is defined as 
\begin{equation}
- \beta F(\beta;\{J_{ij}\}) = \log Z(\beta;\{J_{ij}\}).
\end{equation}
The self-averaging property is valid for other observables, which can be
obtained from the free energy per site. For instance, the magnetization per
site satisfies 
\begin{equation}
m = \frac{1}{N}\sum_{i=1}^N S_i = \langle S_i \rangle \to \left[ \langle S_i
\rangle \right].
\end{equation}
The average through the logarithmic term is still the intractable task. The
replica method is then useful to perform the configurational average for the
free energy. First, we evaluate the averaged power of the partition function
to the natural number $n$ as $\left[ Z^n(\beta;\{J_{ij}\}) \right]$. Then we
consider the analytical continuation of $n$ and take the limit based on the
elementary identity as 
\begin{equation}
\left[ \log Z(\beta;\{J_{ij}\}) \right] = \lim_{n \to 0} \frac{\left[%
Z^n(\beta;\{J_{ij}\})\right] - 1 }{n}.
\end{equation}
You see the replica method everywhere in analyses on spin glasses.

We avoid the complicated details of analytical results in spin glasses in
order to straightforwardly understand the most important parts in this
chapter. However we write down a few things on spin glasses related with our
topics. The peculiar feature in spin glasses is the extraordinary slow
relaxation toward equilibrium state in the low temperature. This is because
the existence of many minima of the free energy on spin glasses revealed by
the mean-field analysis. This fact implies that there are a large number of
the most probable states. The number diverges exponentially as increase of
the number of spins $N$. Unfortunately, this fact has been confirmed by the
mean-field analysis. We have still not given the answer whether the actual
finite-dimensional spin glasses follow the same scenario as that given by
the mean-field analysis or not. The main reason of lack of understanding of
finite dimensional spin glasses is absence of systematic tools to approach
the issue. One of the exceptional methods would be the gauge theory.

\subsection{Gauge theory}

We take the $\pm J$ Ising model as an instance of the spin glass to show the
detailed analysis. For simplicity, we combine the strength interaction with
the inverse temperature as $K = \beta J$. We write the Hamiltonian of the $%
\pm J$ Ising model as 
\begin{equation}
H = - \sum_{\langle ij \rangle} J \tau_{ij}S_iS_j.
\end{equation}
The sign of the interactions follow the distribution function 
\begin{equation}
P(\tau_{ij}) = p \delta(\tau_{ij} - 1) + (1-p)\delta(\tau_{ij} + 1).
\end{equation}
The partition function can be also written as $Z(K;\{\tau_{ij}\})$.

We then define the local transformation by a binary variable $\sigma
_{i}=\pm 1$, called as the gauge transformation, as \cite{HNbook,HN81} 
\begin{eqnarray}
\tau _{ij} &\rightarrow &\sigma _{i}\sigma _{j}\tau _{ij} \\
S_{i} &\rightarrow &\sigma _{i}S_{i}.
\end{eqnarray}%
Notice that this transformation does not change the value of the physical
quantity given by the average over $\tau _{ij}$ and $S_{i}$ since it alters
only the order of the summations. The Hamiltonian can not change its form
after the gauge transformation since the right-hand side is evaluated as 
\begin{equation}
-\sum_{\langle ij\rangle }J\tau _{ij}\sigma _{i}\sigma _{j}\sigma
_{i}S_{i}\sigma _{j}S_{j}=H.
\end{equation}%
As this case, if the physical quantity is invariant under the gauge
transformation (gauge invariant), we can evaluate its exact value even for
finite-dimensional spin glasses. The key point of the analysis by the gauge
transformation is on the form of the distribution function. Before
performing the gauge transformation, the distribution function can take the
following form as 
\begin{equation}
P(\tau _{ij})=\frac{\mathrm{e}^{K_{p}\tau _{ij}}}{2\cosh K_{p}},
\end{equation}%
where $\exp (-2K_{p})=(1-p)/p$. The gauge transformation changes this form,
in a different way from the Hamiltonian, as 
\begin{equation}
P(\tau _{ij})=\frac{\mathrm{e}^{K_{p}\tau _{ij}\sigma _{i}\sigma _{j}}}{%
2\cosh K_{p}}.
\end{equation}

Let us evaluate the internal energy by aid of the gauge transformation here.
The thermal average of the observables $O$ is defined as, by use of the
canonical distribution 
\begin{eqnarray}
\langle O \rangle_{K} &=& \sum_{\{S_i\}} \frac{1}{Z(K;\{\tau_{ij}\})} O
\prod_{\langle ij \rangle } \exp\left( K \tau_{ij}S_iS_j\right).
\end{eqnarray}
Then the thermal average of the Hamiltonian can be written as 
\begin{eqnarray}
\langle H \rangle_{K} &=& \sum_{\{S_i\}} \frac{1}{Z(K;\{\tau_{ij}\})}H
\prod_{\langle ij \rangle } \exp\left( K \tau_{ij}S_iS_j\right)  \label{IE}
\\
&=& - J \frac{d}{dK} \log Z(K;\{\tau_{ij}\}).
\end{eqnarray}
We can use the self-averaging property here since this is given by the
derivative of the free energy, and thus take the configurational average as 
\begin{eqnarray}
\left[ \langle H \rangle_K \right]_{K_p} &=&
\sum_{\{\tau_{ij}\}}\prod_{\langle ij \rangle}\frac{\exp(K_p\tau_{ij})}{%
2\cosh K_p}\times \langle H \rangle_{K},
\end{eqnarray}
where $[\cdots]_{K_p}$ denotes the configurational average with $K_p$. Then
we perform the gauge transformation, which does not change the value of the
internal energy 
\begin{eqnarray}
\left[ \langle H \rangle_{K} \right]_{K_p} &=&
\sum_{\{\tau_{ij}\}}\prod_{\langle ij \rangle}\frac{\exp(K_p\tau_{ij}%
\sigma_i \sigma_j )}{2\cosh K_p}\times \langle H \rangle_{K}.
\end{eqnarray}
Therefore we here take the summation over all the possible configurations of 
$\{\sigma_i\}$ and divide it by $2^N$ (the number of configurations) as 
\begin{eqnarray}
\left[ \langle H \rangle_{K} \right]_{K_p} &=& \frac{1}{2^N}%
\sum_{\{\sigma_{i}\}}\sum_{\{\tau_{ij}\}}\prod_{\langle ij \rangle}\frac{%
\exp(K_p\tau_{ij}\sigma_i \sigma_j )}{2\cosh K_p}\times \langle H
\rangle_{K}.
\end{eqnarray}
We take the summation over $\{\sigma_i\}$ in advance of that over $%
\{\tau_{ij}\}$ and then find the partition function with $K_p$ instead of $K$%
. 
\begin{eqnarray}
\left[ \langle H_{K} \rangle \right]_{K_p} &=& \frac{1}{2^N}%
\sum_{\{\tau_{ij}\}}\frac{Z(K_p;\{\tau_{ij}\})}{(2\cosh K_p)^{N_B}}\times
\langle H \rangle_{K}.
\end{eqnarray}
Going back to Eq. (\ref{IE}), we can delete both of the partition functions
on the denominator and numerator by setting $K_p = K$ as 
\begin{eqnarray}
\left[ \langle H \rangle_{K} \right]_{K} &=& \frac{-J}{2^N(2\cosh K_p)^{N_B}}
\sum_{\{S_{i}\}}\sum_{\{\tau_{ij}\}}\frac{d}{dK}\exp\left( K
\tau_{ij}S_iS_j\right)  \nonumber \\
& =& -N_B \tanh K.
\end{eqnarray}
Similarly, we can evaluate the rigorous upper bound on the specific heat as
well as the restriction on the structure of the phase diagram. The condition 
$K_p = K$ defines the special subspace in which we can perform the exact
analysis even for finite-dimensional spin glasses. This subspace is called
as the Nishimori line \cite{HNbook,HN81}. On this subspace, we can reveal
several rigorous properties on the structure of the phase diagram even for
spin glasses by relatively simple calculations. For instance, let us
consider to evaluate the local magnetization $\langle S_i \rangle$ by the
gauge transformation. The local magnetization identifies the existence of
the ferromagnetic order and its definition is 
\begin{equation}
\langle S_i \rangle_{K} = \sum_{\{S_{i}\}}S_i \prod_{\langle ij \rangle}%
\frac{\exp(K\tau_{ij}S_iS_j)}{Z(K;\{\tau_{ij}\})}.
\end{equation}

\begin{figure}[tbp]
\begin{center}
\includegraphics[width=80mm]{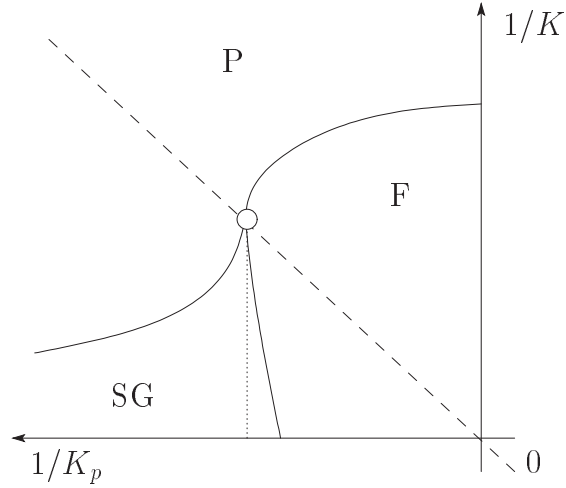}
\end{center}
\caption{Nishimori line and the typical phase diagram of the $\pm J$ Ising
model. The vertical axis is the temperature, since $K$ can be regarded as
the inverse temperature. The horizontal axis expresses the density of the
ferromagnetic interactions in terms of $K_{p}$. The ferromagnetic,
paramagnetic, and spin-glass phases are denoted by \textquotedblleft F",
\textquotedblleft P" and \textquotedblleft SG", respectively. The solid
curve represent the phase boundary. The dashed curve describes the Nishimori
line. The multicritical point is located at the most left-hand side. }
\label{PG3D}
\end{figure}

Let us consider to evaluate its configurational average 
\begin{equation}
\left[ \langle S_{i}\rangle _{K}\right] _{K_{p}}=\sum_{\{\tau
_{ij}\}}\prod_{\langle ij\rangle }\frac{\exp (K_{p}\tau _{ij})}{2\cosh K_{p}}%
\times \langle S_{i}\rangle _{K}.
\end{equation}%
After the gauge transformation, we obtain 
\begin{equation}
\left[ \langle S_{i}\rangle _{K}\right] _{K_{p}}=\sum_{\{\tau _{ij}\}}\sigma
_{i}\prod_{\langle ij\rangle }\frac{\exp (K_{p}\tau _{ij}\sigma _{i}\sigma
_{j})}{2\cosh K_{p}}\times \langle S_{i}\rangle _{K}.
\end{equation}%
By summing over all the possible configurations of the gauge variables and
dividing the resulting equality by $2^{N}$, we reach 
\begin{equation}
\left[ \langle S_{i}\rangle _{K}\right] _{K_{p}}=\sum_{\{\tau _{ij}\}}\frac{%
Z(K_{p};\tau _{ij})}{2^{N}\left( 2\cosh K_{p}\right) ^{N}}\times \langle
\sigma _{i}\rangle _{K_{p}}\times \langle S_{i}\rangle _{K}.
\end{equation}%
On the other hand, let us evaluate the configurational average of the
product of the correlation functions with different temperatures as 
\begin{equation}
\left[ \langle \sigma _{i}\rangle _{K_{p}}\langle S_{i}\rangle _{K}\right]
_{K_{p}}=\sum_{\{\tau _{ij}\}}\prod_{\langle ij\rangle }\frac{\exp
(K_{p}\tau _{ij})}{2\cosh K_{p}}\times \langle \sigma _{i}\rangle
_{K_{p}}\langle S_{i}\rangle _{K}.
\end{equation}%
On this equality, the gauge transformation only changes the sign of the
interactions as $\tau _{ij}\rightarrow \tau _{ij}\sigma _{i}^{\prime }\sigma
_{j}^{\prime }$ in the exponential function. We find the following equality
by summing over $\{\sigma _{i}^{\prime }\}$ and multiplying $1/2^{N}$ 
\begin{equation}
\left[ \langle S_{i}\rangle _{K}\right] _{K_{p}}=\left[ \langle \sigma
_{i}\rangle _{K_{p}}\langle S_{i}\rangle _{K}\right] _{K_{p}}.
\end{equation}%
Let us discuss the structure of the phase diagram of the $\pm J$ Ising model
by use of this equality. Setting $K=K_{p}$, we obtain 
\begin{equation}
\left[ \langle S_{i}\rangle _{K_{p}}\right] _{K_{p}}=\left[ \langle
S_{i}\rangle _{K_{p}}^{2}\right] _{K_{p}}.  \label{QMNL}
\end{equation}%
The quantity on the right-hand side is the order parameter of the spin-glass
phase, called as the spin-glass parameter. In the spin-glass phase, the
directions of the spins orient are random in space. Thus $m$ should be zero
and it implies 
\begin{equation}
m=\frac{1}{N}\sum_{i}S_{i}=\langle S_{i}\rangle \rightarrow \left[ \langle
S_{i}\rangle _{K_{p}}\right] _{K_{p}}\rightarrow 0.
\end{equation}%
On the other hand, if the spins are frozen, their square should be non-zero.
Thus 
\begin{equation}
q=\left[ \langle S_{i}\rangle _{K_{p}}^{2}\right] _{K_{p}}\neq 0.
\end{equation}%
Therefore, on the Nishimori line, we prove that the spin-glass phase does
not exist since the important fact $q=m$ is given by Eq. (\ref{QMNL}). In
addition, if we take the absolute value of the magnetization away from the
Nishimori line, we find 
\begin{equation}
|\left[ \langle S_{i}\rangle _{K}\right] _{K_{p}}|\leq |\left[ \langle
S_{i}\rangle _{K}\right] _{K_{p}}|\times |\left[ \langle S_{i}\rangle
_{K_{p}}\right] _{K_{p}}|\leq |\left[ \langle S_{i}\rangle _{K_{p}}\right]
_{K_{p}}|.
\end{equation}%
This inequality states that the magnetization takes the largest value on the
Nishimori line along the vertical line of $K_{p}$. It implies that a special
critical point, multicritical point, is located at the most left-hand side
of the phase diagram as in Fig. \ref{PG3D}. 

In the following sections, we will find several problems on spin glasses,
which are lying between statistical mechanics and quantum information
processing. Although it is not enough to understand the whole properties on
spin glasses, let us first go to these fascinating parts. After showing the
interesting connections, we will strive the specific problems related with
the spin glasses we must solve then.

\section{Quantum error correction: surface code}

In the first part of this chapter, we describe a technique for the quantum
error correction. In order to protect the vulnerable quantum state from
decoherence, we consider a method for circumvention of the effects from
decoherence. It is very important to perform the quantum information
processing. We here show an elaborated technique by use of the property of
the topology. Our approach, in short terms, is based on encoding a few
``logical" qubits in a particular state, which is not disturbed directly by
noise of the ``physical" qubits. Here we call the quantum state describing
superposition of the binary state $0$ and $1$ (in terms of informatics,
bits) as qubits This strategy is analogous to the classical counterpart, in
which we introduce some redundancy to restore the original state.

\subsection{Error model}

The quantum state of the single qubit is written as 
\begin{equation}
|\Psi \rangle = \alpha |0 \rangle + \beta |1 \rangle.
\end{equation}
Here the above coefficients follow $|\alpha|^2 + |\beta|^2 = 1$. We can
write all the changes on the binary Hilbert space by combination of the
identity operator and the Pauli operators $X$, $Y$ and $Z$. 
\begin{equation}
X =\left( 
\begin{array}{cc}
0 & 1 \\ 
1 & 0%
\end{array}
\right), \quad Y =\left( 
\begin{array}{cc}
0 & \mathrm{i} \\ 
-\mathrm{i} & 0%
\end{array}
\right), \quad Z=\left( 
\begin{array}{cc}
1 & 0 \\ 
0 & -1%
\end{array}
\right).
\end{equation}
By this mean, we express the error by the action of the Pauli operator. The
action of $X$ represents a phase error, and that of $Z$ expresses a flip
error. In addition, $Y$ is a multiple error as $Y= \mathrm{i}X Z$.

Assuming the error will occur in a stochastic manner, let us define a noise
model where the qubit gets errors as 
\begin{equation}
\rho \to p_I\rho + \left(p_XX\rho X + p_YY\rho Y + p_ZZ\rho Z\right).
\label{error}
\end{equation}
Although we can deal with any cases of $p_I,p_X,p_Y$, and $p_Z$, Let us
start from the simple case with uncorrelated between the flip and phase
errors as $p_I= (1-p)^2$ and $p_X = p_Z = p$, while $p_Y=p^2$, where $0 \le
p \le 1$. In this case, we can independently treat two of the errors. We
must construct an ingenious procedure to recover the damaged qubits, that is
the quantum error correction. Once an error occurs on the single qubit
system, we can not remove the error, since we do not know the original
state. The single qubits is too weak to save the specific information once
error occurs. Therefore we need to prepare many qubits against errors as the
first strategy. The first stage of the quantum error correction is thus to
construct the many-body quantum system to store several bits of the
information. That is analogous with the concept of redundancy in the classical
information.

\subsection{Surface code}

\begin{figure}[tbp]
\begin{center}
\includegraphics[width=100mm]{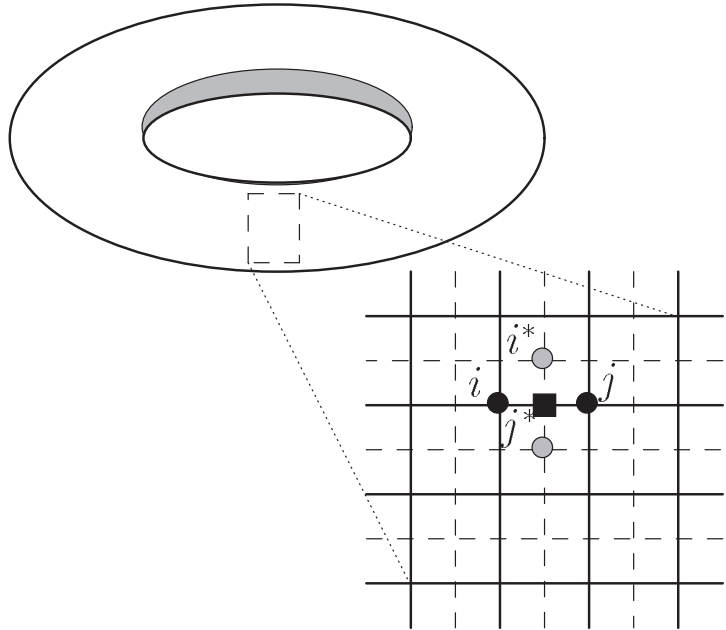}
\end{center}
\caption{Square lattice (say original) and its dual on a torus. The location
of the qubit is the center of the edge $(ij)$ (or $(i^*j^*)$) denoted by the
black square. }
\label{Sq}
\end{figure}

\begin{figure}[tbp]
\begin{center}
\includegraphics[width=100mm]{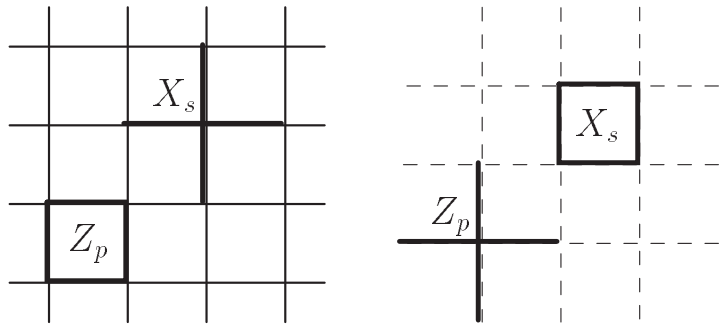}
\end{center}
\caption{Star and plaquette operators. }
\label{CO}
\end{figure}

\begin{figure}[tbp]
\begin{center}
\includegraphics[width=100mm]{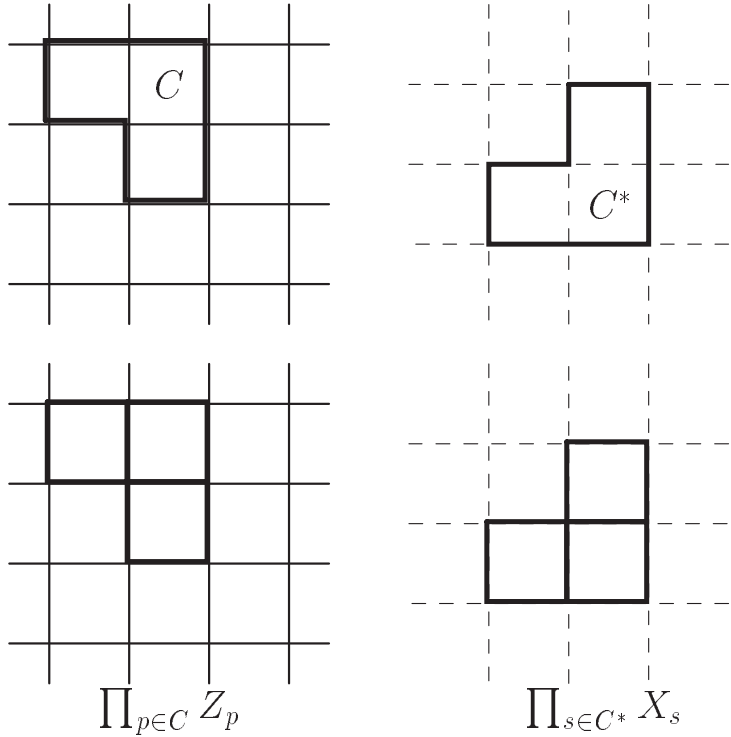}
\end{center}
\caption{Contractible loops $C$ and $C^*$ and their expressions by the check
operators $X_s$ and $Z_p$. }
\label{UL}
\end{figure}

Let us construct an array of qubits on a torus by setting qubits on each
edge $(ij)$ of the square lattice embedded on a torus (genus $1$). Here we
make two ways to describe the square lattice, while being unchanged of the
location of the qubits, as in Fig. \ref{Sq}. 
The original square lattice is convenient to explain how to detect and
remove the flip errors and the other (dual) is for the phase errors. The
embedded qubits (say physical qubits) on all the edges actually suffer from
disturbance by noisy environment. We do not directly use the quantum state
of these physical qubits to encode the information. Instead, we make the
particular quantum state, which is stable against direct effects due to the
errors. In order to construct such a fault-tolerant subspace, called as the
codespace, for encoding the information, we define two of the check
operators. One is the star operator for each site $s$ as 
\begin{equation}
X_s = \otimes_{(ij) \in s}X_{(ij)},
\end{equation}
and the other is the plaquette operator for each plaquette $p$ 
\begin{equation}
Z_p = \otimes_{(ij) \in p}Z_{(ij)},
\end{equation}
where the product consists of four edges adjacent to each site and each
plaquette as depicted in Fig. \ref{CO}. 
All of the star operators $X_s$ and the plaquette operators $Z_p$ commute
and are thus simultaneously diagonalizable. The codespace to encode the
specific information consists of their simultaneous $+1$ eigenstates of all
the check operators as $X_s|\Psi_c\rangle = +1 | \Psi_c \rangle$ for any $s$
and $Z_p|\Psi_c \rangle = +1 | \Psi_c \rangle$ for any $p$. The site on the
original square lattice is replaced by the plaquette on the dual one, and
vice versa. The star operators are thus found to consist of the unit loops
on the ``dual" square lattices. Let us consider contractible loops on the
surface of the torus denoted by $C$ and $C^*$ on the original and dual
lattices, respectively. The action $C^*$ can be expressed by the product of $%
X_s$ as in Fig. \ref{UL}. 
Here notice that the double action of the same Pauli operator becomes unity.
On the other hand, the plaquette operators conform the unit loop on the
original square lattice. Thus the action represented by any contractible
loops on the original lattice $C$ is given by the product of $Z_p$ as in
Fig. \ref{UL}. The action described by $C$ and $C^*$ acts trivially on the
codespace, since the codespace is the eigenspace of the star and plaquette
operators. In other words, the compatible-loop actions by the Pauli operator 
$Z$ on the original lattice and $X$ on the dual one can not alter the
particular quantum state encoded on the torus.

We here show the explicit form of the encoded quantum state below. Let us
prepare a uniform linear-combination state of all the possible contractible
loops with the products of $X_s$ and $Z_p$ as 
\begin{eqnarray}
| \Psi_c \rangle &=& \left( 1 + \sum_s X_s + \sum_{s_1,s_2} X_{s_1,s_2} +
\cdots \right)  \nonumber \\
& &\quad \times \left( 1 + \sum_p Z_p + \sum_{p_1,p_2} Z_{p_1}Z_{p_2} +
\cdots \right) |\Psi_0 \rangle,
\end{eqnarray}
where $|\Psi_0\rangle$ is the vacuum state of the physical qubits. Then we
can easily confirm that $X_s| \Psi_c \rangle = | \Psi_c \rangle$ and $Z_p |
\Psi_c \rangle = | \Psi_c \rangle$.

\begin{figure}[tbp]
\begin{center}
\includegraphics[width=100mm]{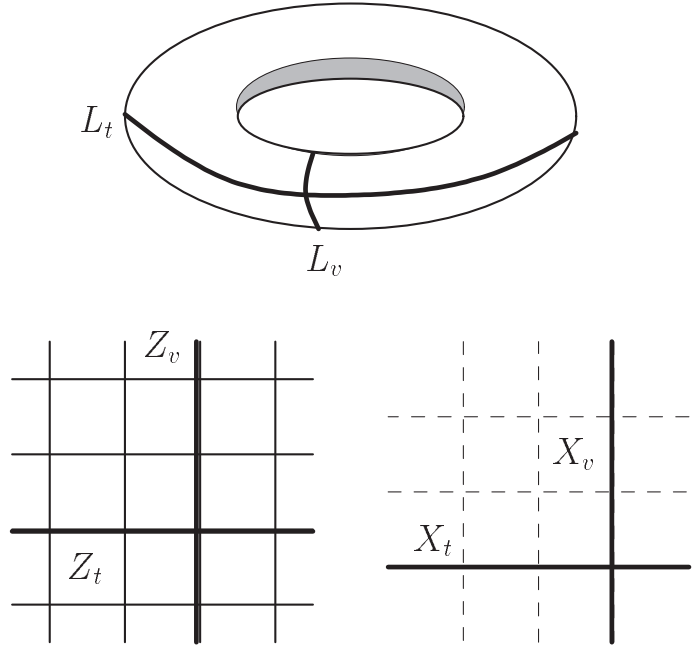}
\end{center}
\caption{Non-contractible loops $L_v$ and $L_t$ and the logical operators. }
\label{LO}
\end{figure}

On the other hand, any non-contractible loop, winding around the torus, of
the Pauli operators of $X$ and $Z$ can map the codespace to itself in a
nontrivial manner, since the possible combinations of the contractible loops
never constitute any non-contractible loops, while such an operator can
commute with any star and plaquette operators. If we set $L \times L$
lattice on a torus, we have $2L^2$ physical qubits and $2(L^2-1)$ check
operators. The remaining degrees of freedom of $2$ implies existence of two
non-contractible loops, winding around the hole of the torus $L_v$ and
winding around the body of the torus $L_t$ as depicted in Fig. \ref{LO}. 
Let us express these non-contractible loops on the torus in terms of the
products of Pauli operators as 
\begin{eqnarray}
\bar{X}_v &=& \prod_{(ij) \in L_v}X_{(ij)} \\
\bar{X}_t &=& \prod_{(ij) \in L_t}X_{(ij)} \\
\bar{Z}_v &=& \prod_{(ij) \in L^*_v}Z_{(ij)} \\
\bar{Z}_t &=& \prod_{(ij) \in L^*_t}Z_{(ij)}.
\end{eqnarray}
They are termed as logical operators. Here we use the asterisk denoting the
non-contractible loop on the dual lattice. The combinations of
non-contractible loops yield $2^4 = 16$ different homology classes embedded
in the original and dual square lattices on the single torus. The elementary
manipulation confirms that the logical operators can form Pauli algebra of
two effective qubits encoded in the topological degrees of freedom on the
torus as $[\bar{Z}_v,\bar{Z}_t]=[\bar{X}_v,\bar{X}_t]=0$, while $\bar{X}_v%
\bar{Z}_t = -\bar{Z}_t\bar{X}_v$ and $\bar{X}_t\bar{Z}_v = -\bar{Z}_v\bar{X}%
_t$. Thus, by use of these algebras by the non-contractible loops on the
torus, we can prepare a $4 \times 4$ Hilbert space, say two logical qubits,
explicitly given by $|\Psi_{k_1,k_2,k_3,k_4} \rangle =
Z^{k_1}_vZ^{k_2}_tX^{k_3}_vX^{k_4}_t|\Psi_c \rangle$, where $k_i$ is the
number taking $0$ and $1$ to distinguish the basis of the logical qubits.
The logical qubits is in general written as 
\begin{equation}
|\Psi \rangle = \sum_{k_1,k_2,k_3,k_4} A_{k_1,k_2,k_3,k_4}
|\Psi_{k_1,k_2,k_3,k_4} \rangle,
\end{equation}
where $A_{k_1,k_2,k_3,k_4}$ is the coefficient following $%
\sum_{k_1,k_2,k_3,k_4}|A_{k_1,k_2,k_3,k_4}|^2 = 1$.

We can indeed prepare the particular quantum state from many redundant
qubits as shown above. Our next interest should be how this quantum subspace
is stable against the errors on the physical qubits.

\subsubsection{Check operators and error syndrome}

\begin{figure}[tbp]
\begin{center}
\includegraphics[width=100mm]{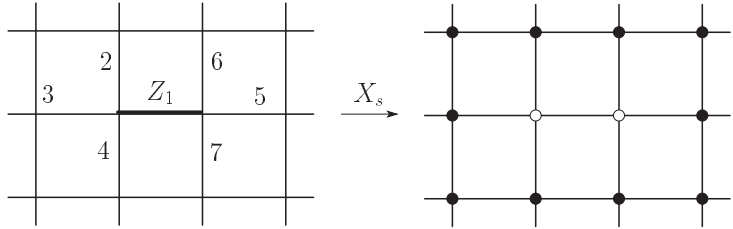}
\end{center}
\caption{Detection of the error by the check operator. The left panel shows
the original error chain $E$ on the edge $1$. The right panel describes the
outputs of the star operators, which describe the endpoints of the error $%
\partial E$. The white circles denote nontrivial outputs $-1$ and the black
ones represent $+1$. }
\label{ES}
\end{figure}

The effect coming from the decoherence, error, can occur everywhere on the
torus. Let us assume that the error can be individually independently
generated on each physical qubit on the torus through the channel model
defined in Eq. (\ref{error}). The errors $Z_{(ij)}$ and $X_{(ij)}$ can be
described as, in general, non-closed error chains $E$ and $E^*$ on both of
the original and dual lattices. We should detect the location of the errors
to circumvent their effects. The star and plaquette operators can also play
a roll to generate the partial information of the errors on the physical
qubits. Thus these are often called the check operators. The endpoints of
the error chains $\partial E$ and $\partial E^*$ can be detected by
applications of star and plaquette operators due to anti-commutation of
error with adjacent operators as follows. For instance, let us consider the
local error by the action of $Z$ on the original square lattice as in Fig. %
\ref{ES}. 
The action of the error can be written as $Z_1|\Psi \rangle $. If we apply
the star operator around $Z_1$ on the square lattice, we obtain a pair of
the outputs with the negative value $-1$ as 
\begin{eqnarray}
X_1X_2X_3X_4 Z_1|\Psi \rangle = - Z_1 X_1X_2X_3X_4|\Psi \rangle = - Z_1|\Psi
\rangle \\
X_1X_5X_6X_7 Z_1|\Psi \rangle = - Z_1 X_1X_5X_6X_7|\Psi \rangle = - Z_1|\Psi
\rangle,
\end{eqnarray}
where we use anti commutation as $Z_1X_1 = - X_1Z_1$. On the other hand, the
flip error by $X$ can be detected by the plaquette operator on the dual
lattice. At least, we can identify the location of the endpoints on the
error chains denoted as $\partial E$ and $\partial E^*$.

We do not know the exact shape of the error chains $E$ and $E^*$. We must
remove the effect of the errors on the torus only by use of the outputs of
the check operators, error syndrome, $\partial E$ and $\partial E^*$. Thus
we must infer the original state from the damaged one with negative outputs
on several sites and plaquettes. To remove the damage signaled by the
negative eigenvalues, we consider to apply additional actions of the product
of the Pauli operators. This procedure is to simply connect two endpoints of
the error chains we can know. The additional actions with the original error
chains can conform closed loops. The resulting effects are trivial on the
codespace. However, if nontrivial, we can no longer restore the original
quantum state.

In order to make the resulting effect trivial, we have to choose additional
actions while inferring the same homology class as the logical operators
encoding the original information only with the knowledge of the endpoints.
The best way (optimal procedure) is to find out the most probable homology
class. It reads, if we define $P(\bar{E},\bar{E}^*|\partial E,\partial E^*)$
as the probability of the inferred homology class $\bar{E}$ and $\bar{E}^*$
conditioned on $\partial E$ and $\partial E^*$, 
\begin{equation}
\mathrm{max}_{\bar{E},\bar{E}^*} P(\bar{E},\bar{E}^*|\partial E,\partial
E^*),
\end{equation}
where $\bar{E}$ and $\bar{E}^*$ denote the possible homology class of the
error chains. Observant readers begin to recognize an existence of the
limitation of the above procedure. First, increase of the error probability $%
p$ might allow the possibility of the non-contractible error chains wounding
the torus. If we prepare sufficient large torus, this is not expected to
occur as $\sim p^L \to 0$. As this instance, the system size $L$ becomes
larger, tolerance of the surface code against the error can be enhanced.
However, if $p$ is not small, we can not always determine the same homology
class with the error chains among several candidates, since we tend to infer
the wrong homology class due to the existence of long and many error chains.
The connections among the original error chains and the additional actions
would conform a long loop wounding the torus if $p$ exceeds a threshold.
Such an accuracy threshold actually exists and can be determined by special
technique developed in the theory on the phase transition in statistical
mechanics. In practice, the ideal error correction by inferring the most
probable homology class of the error chains should be too harmful due to
computational costs. We need to take some approximate technique in a
moderate time. However, if we can, it is important to estimate the optimal
error threshold in order to reveal the theoretical limitation of the surface
code.

In the following, we construct the statistical-mechanical model to pave the
way to analytically estimate the optimal accuracy threshold.

\subsubsection{Probability of error chains}

Notice that our task is to elucidate the macroscopic property in the error
correcting procedure, not actual many-body quantum system. The outputs of
the measurements of the check operators and the procedure of the application
of additional actions based on the inferred homology class are classical.
Although we deal with the problem on the quantum error correction, all the
methods we employ here may not be quantum.

The error on each qubits occur in the probabilistic manner. Therefore let us
construct a probabilistic model of the quantum error correction on the
torus. In order to deal with the asymptotic behavior of the error pattern
and inferred chains in the infinite-number limit, we use the most suited
technique, statistical mechanics. Statistical mechanics starts from
constructing a probabilistic model in the microscopic level. Following the
concept of the large deviation principle, we take an infinite number limit
of the system size in order to predict the deterministic macroscopic
behavior.

Now we take the simple case with uncorrelated errors. We can separately deal
with the flip and phase errors. For simplicity, hereafter we consider only
the phase errors given by the actions of $Z$ on the original square lattice.

From the knowledge of endpoints $\partial E$, error syndrome, we have to
infer the most likely homology class of the error chains, while considering
any reasonable choices.

\begin{figure}[tbp]
\begin{center}
\includegraphics[width=100mm]{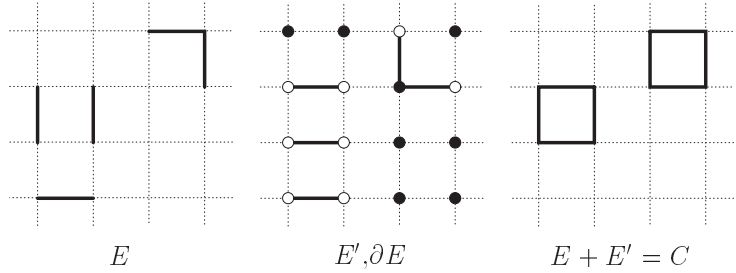}
\end{center}
\caption{Reasonable connections $E^{\prime }$ inferred from $\partial E$.
For transparency, the square lattice is depicted by the dotted lines. }
\label{RC}
\end{figure}

For instance, let us consider the reasonable connections $E^{\prime }$
between each pair of the endpoints $\partial E$ as in Fig. \ref{RC}. 
As a result, we can make contractible loops denoted by $C$ on the square
lattice. Then the reasonable chains $E^{\prime }$ inferred from $\partial E$
are in an equivalent class with the error chains. However there are several
cases to infer non-equivalent class with the error chains as in Fig. \ref%
{fRC}. 
\begin{figure}[tbp]
\begin{center}
\includegraphics[width=100mm]{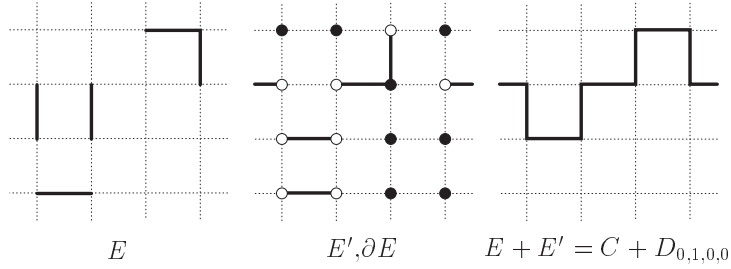}
\end{center}
\caption{Failure of inference of the equivalent class with $E$. The wounding
chain by the product of $Z$ yields $Z_t$ denoted by $(k_1,k_2,k_3,k_4) =
(0,1,0,0).$ }
\label{fRC}
\end{figure}

By use of the Bayes theorem, we consider to evaluate the probability to
infer the equivalent homology class with the original error chains
conditioned on those endpoints as follows 
\begin{equation}
P(\bar{E}|\partial E) = \frac{P(\partial E|\bar{E})P(\bar{E})}{%
\sum_{k_1,k_2,k_3,k_4}P(\partial E|\bar{E}+D_{k_1,k_2,k_3,k_4})P(\bar{E}%
+D_{k_1,k_2,k_3,k_4})},
\end{equation}
where $D_{k_1,k_2,k_3,k_4}$ denotes the logical operator. The equivalent
class is given simply by $D_{0,0,0,0} = \phi$. If we infer the different
class denoted by $(k_1,k_2,k_3,k_4) \neq (0,0,0,0)$, the error correction
fails. We assume that the prior probability for the homology classes is
uniform, namely $P(\partial E) = 1/16$, since we encode various combinations
of the different homology classes. Our task is to consider $P(\partial E|%
\bar{E})$, which is the probability to generate the endpoints of the
specific error chains. Let us first evaluate the probability for the
specific error chains. The error chains are generated following the
independently identical distribution as 
\begin{eqnarray}
P(E) &=& \prod_{\langle ij \rangle }p^{\frac{1+\tau^E_{ij}}{2}}(1-p)^{\frac{%
1-\tau^E_{ij}}{2}}  \nonumber \\
&=& \left\{p(1-p)\right\}^{\frac{N_B}{2}}\prod_{\langle ij \rangle }\left(%
\frac{p}{1-p}\right)^{\frac{\tau^E_{ij}}{2}},
\end{eqnarray}
where $\tau^{E}_{ij}$ represents the error chains and takes $\pm 1$ ($%
\tau^{E}_{ij}<0$, when $(ij)\in E$). Here we introduce the expression of the
exponential form as in the case for the spin glass and obtain 
\begin{equation}
P(E) = \prod_{\langle ij \rangle}\frac{\exp(K_p\tau^E_{ij})}{2\cosh K_p}.
\end{equation}
Notice that $K_p$ is minus of the original definition in the spin glass.
Then, in order to evaluate $P(\partial E|\bar{E})$, we sum over all the
possible configurations $E^{\prime }= E + C$ of the error chains sharing the
endpoints. We reach 
\begin{equation}
P(\partial E | \bar{E}) \propto \sum_{C}\prod_{\langle ij\rangle}\exp(K_p
\tau^{E}_{ij}\tau^{C}_{ij}),
\end{equation}
where the summation is taken over all the possibilities of $C$ and the
product is over all the edges. We use the same indicators as $\tau_{ij}^{E}$
for $C$.

This probability tells us how to proceed the error correction. We simply
perform the connection of the endpoints in the stochastic manner. If we know 
$p$ in advance, it is better to set the same value. Otherwise we have 
\begin{equation}
P_K(\partial E | \bar{E}) \propto \sum_{C}\prod_{\langle ij\rangle}\exp(K
\tau^{E}_{ij}\tau^{C}_{ij}).
\end{equation}
The parameter $K$ stands for the importance/preference to choose the
reasonable choice in the connection of the endpoints. For instance, in the
case with $K \to \infty$, $E^{\prime }$ with the minimum length are
preferred. The performance of the error correction depends on the value of $K
$.

The loop constraints $\prod_{(ij)}\tau_{ij}^C=1$ allow to use another
expression by the Ising variables $\tau_{(ij)}^C=S_iS_j$. By use of this
expression, we can find that $P(\bar{E})$ is written by the partition
function of the $\pm J$ Ising model as 
\begin{equation}
P_K(\partial E | \bar{E}) \propto Z(K;\{\tau_{ij}\}) =
\sum_{\{S_i\}}\prod_{\langle ij \rangle}\mathrm{e}^{K\tau^{E}_{ij} S_iS_j},
\label{PF1}
\end{equation}
where $\tau^{E}_{ij}$ is the sign of the random coupling in context of spin
glasses. Each of the random couplings follows the distribution function of
the error chains $P(E)$. Notice that, in this case, $p$ denotes the density
of the antiferromagnetic interactions. Therefore the problem to identify the
error threshold of the surface code is reduced into evaluating the quantity
related with the partition function of the simple spin-glass model, the $\pm J
$ Ising model on the square lattice. If we increase $p$, the spin-glass
system will be not rigid against thermal fluctuations, since the
ferromagnetic order decays. Similarly, in the surface code, increase of $p$
is expected to lead the error correction to be unfeasible. These analogous
implies the existence of a fascinating relationship between the phase
transition in the spin glass and the error threshold in the surface code.

\begin{figure}[tbp]
\begin{center}
\includegraphics[width=80mm]{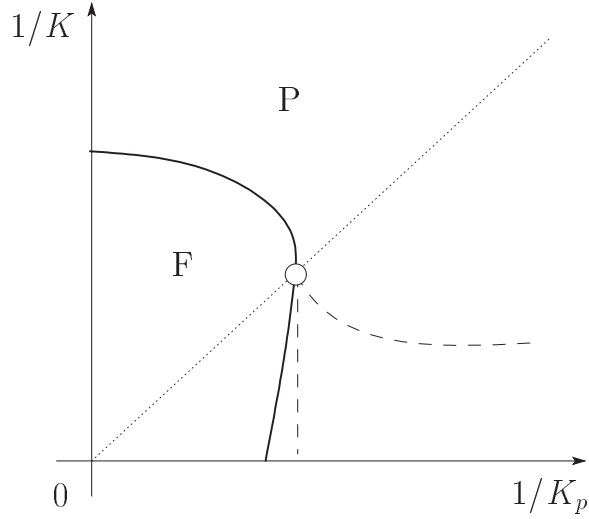}
\end{center}
\caption{Phase diagram of the $\pm J$ Ising model. The figure is depicted by
the same symbols in Fig. \protect\ref{PG3D}. In the two-dimensional case,
the absence of the spin-glass phase has been partially proved and supported
by numerical simulations. \protect\cite{AbSG}. }
\label{PG}
\end{figure}

In order to more clearly see this connection between statistical mechanics
and quantum error correction, we show a schematic picture of the phase
diagram for the $\pm J$ Ising model on the square lattice as in Fig. \ref{PG}%
. 
In the low-temperature region with a relatively small $p$, the ferromagnetic
ordered phase can be observed. It means that the order of the Ising spins
exists and suppresses the fluctuation of the domain walls (boundaries
between different signed spins). Let us consider to evaluate $P_K(\bar{E}%
|\partial E)$ in the ferromagnetic phase. We rewrite the probability as, in
terms of the free energy of the spin glass, 
\begin{equation}
P_K(\bar{E}|\partial E) = \frac{1}{1+\sum_{\{k_i\}}\exp(-K \Delta
F_{k_1,k_2,k_3,k_4})}.
\end{equation}
Here the difference of the free energy $\Delta F_{k_1,k_2,k_3,k_4}$ is given
as 
\begin{equation}
\Delta F_{k_1,k_2,k_3,k_4} = F(K;\{\tau_{ij}\}) -
F(K;\{\tau_{ij}\tau^{D_{k_1,k_2,k_3,k_4}}_{ij}\}),
\end{equation}
where we use $K$ instead of $\beta$ as 
\begin{equation}
F(K;\{\tau_{ij}\tau^{D_{k_1,k_2,k_3,k_4}}_{ij}\}) = -\frac{1}{K}\log
Z(K;\{\tau_{ij}\tau^{D_{k_1,k_2,k_3,k_4}}_{ij}\}).
\end{equation}
In context of statistical mechanics, the logical operator $%
D_{k_1,k_2,k_3,k_4}$ is interpreted as the induction of the
antiferromagnetic interactions along the non-contractible loop as $%
\{\tau_{ij}\tau^{D_{k_1,k_2,k_3,k_4}}_{ij}\}$. In the ferromagnetic phase,
such a long boundary of antiferromagnetic interactions yields a crevice with
the same scale as the inducted loop in the ordered spins. Therefore the free
energy difference should become $\sim \mathcal{O}(L)$. As a result, the
probability to infer the equivalent homology class with the original error
chains is asymptotically, in the limit of $L \to \infty$ 
\begin{equation}
P_K(\bar{E}|\partial E) \to 1.
\end{equation}
This means that the error correction is feasible.

On the other hand, in a high temperature and for not a small $p$, namely the paramagnetic
phase, the order of the spins is destroyed. 
The vast number of islands of
the different-signed spins can exist and the boundaries fluctuate. The
free-energy difference becomes zero, and thus we reach. 
\begin{equation}
P_K(\bar{E}|\partial E) \to 1/16.
\end{equation}
This means that the error correction is infeasible. Therefore the locations
of the critical points of the spin-glass model identify the error thresholds
of the surface code. In particular, the optimal threshold is located on the
special critical point along the Nishimori line, namely multicritical point 
\cite{HNbook,HN81}, since it is at the most right side in the phase diagram.
In other words, if we know the precise value of $p$, it is easier to correct
the errors in the logical qubits.

\subsection{Analyses on accuracy thresholds for surface code}
In general, the problem on finite-dimensional spin glasses is intractable.
However recent development in theory of this realm enables us to estimate a
precise location of the critical point in several spin glasses. In the
following section, we introduce such a specialized theory in detail.

\subsubsection{Duality analysis: simple case}

The situation that no systematic analytical methods attacking the problems on the
critical phenomena in finite-dimensional spin glasses have been changed since a recent
development in the spin-glass theory.
It enables us to estimate the precise
value of the special critical point especially on the Nishimori line, which
corresponds to the optimal error threshold \cite{NN,MNN,ONB,Ohzeki}. The
method as shown below is based on the duality, which can identify the
location of the critical point especially on two-dimensional spin systems 
\cite{KW,WuWang}. Let us review the simple case of the Ising model.

The duality is a symmetry argument by considering the low and
high-temperature expansions of the partition function. As a result, we can
obtain a simple relation between two different temperatures through the
partition function as 
\begin{equation}
Z(K) = \Lambda Z(K^*),
\end{equation}
where $K^*$ denotes the dual coupling constant related with the original one 
$K$. The coefficient $\lambda$ will be an important quantity below.

We here deal with the case for the Ising model on the square lattice, whose
partition function is given as, through the Hamiltonian (\ref{IsingH}), 
\begin{equation}
Z(K) = \sum_{\{S_i\}}\prod_{\langle ij \rangle}\exp(KS_iS_j),
\end{equation}
where the product is taken over all the nearest neighboring pairs on the
square lattice. The original formulation of the duality analysis is based on
a relatively painful calculation \cite{KW}. We here show a much simpler
version given by a simple Fourier transformation for the local part of the
Boltzmann factor \cite{WuWang}.

\begin{figure}[tbp]
\begin{center}
\includegraphics[width=80mm]{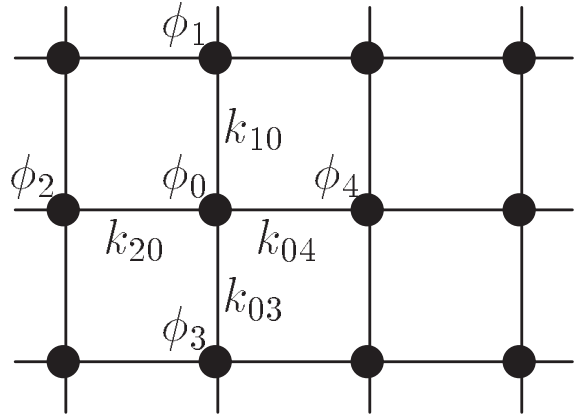}
\end{center}
\caption{Description of the terms appearing in Eq. (\protect\ref{Kd}) }
\label{DS}
\end{figure}

We define the edge Boltzmann factor as 
\begin{equation}
x_{\phi}(K) = \exp(K\cos \pi \phi),
\end{equation}
where we use another binary variable as $\phi = 0$ and $1$, instead of the
Ising spin. In addition, we apply the binary Fourier transformation to this
quantity, called as the dual edge Boltzmann factor, 
\begin{equation}
x^*_{k}(K) = \frac{1}{\sqrt{2}}\sum_{\phi=0,1}x_{\phi}\exp(\mathrm{i}\pi k
\phi) = \frac{1}{\sqrt{2}}\left( \mathrm{e}^{K} + \mathrm{e}^{-K}\cos \pi k
\right).
\end{equation}
Then we find that the partition function can be written in two ways by both
of the edge Boltzmann factors. First, the partition function is simply
expressed by the original Boltzmann factor as 
\begin{equation}
Z(K) = \sum_{\{\phi_i\}}\prod_{\langle ij \rangle}x_{\phi_{ij}}(K),
\end{equation}
where $\phi_{ij} = \phi_i - \phi_j$. The difference between $\phi_i$ and $%
\phi_j$ is taken in order from left to right and from top to bottom.
Inserting the inverse Fourier transformation of the dual edge Boltzmann
factor yields another expression of the partition function 
\begin{equation}
Z(K) = \left( \frac{1}{\sqrt{2}}\right)^{N_B}
\sum_{\{\phi_i\}}\sum_{\{k_{ij}\}}\prod_{\langle ij \rangle}x^*_{k_{ij}}(K)%
\mathrm{e}^{\mathrm{i}k_{ij}(\phi_i - \phi_j)}.
\end{equation}
Here let us perform the summation over $\{\phi_i\}$ by considering the
adjacent edges. We concentrate on the particular site $0$ and then find that
the exponential term can be factorized and the summation can be taken
independently as 
\begin{equation}
\sum_{\phi_0=0,1} \mathrm{e}^{\mathrm{i}(-k_{10} - k_{20} + k_{03} +
k_{04})\phi_0 } = 2 \delta(k_{10} + k_{20} - k_{03} -k_{04}\equiv 0~(\mathrm{%
mod}~2)).  \label{Kd}
\end{equation}
We describe adjacent edges to the particular site $0$ as in Fig. \ref{DS}. 
In order to remove the terms of the Kronecker's delta functions, we employ
another binary variable as 
\begin{eqnarray}
k_{10} &=& \phi^*_1 - \phi^*_2 \\
k_{20} &=& \phi^*_2 - \phi^*_3 \\
k_{03} &=& \phi^*_4 - \phi^*_3 \\
k_{04} &=& \phi^*_1 - \phi^*_4.
\end{eqnarray}
We set these new variables $\{\phi^*_i\}$ on each plaquette on the square
lattice, namely each site on the dual one. The resultant expression of the
partition function is 
\begin{equation}
Z(K) = \sum_{\{\phi^*_i\}}\prod_{\langle ij \rangle}x^*_{\phi^*_{ij}}(K),
\end{equation}
where $\phi^*_{ij} \equiv \phi^*_i - \phi^*_j$. This fact leads us to the
double expressions with the different arguments as 
\begin{equation}
Z(x_0(K),x_1(K)) = Z(x^*_0(K),x^*_1(K)).
\end{equation}
To reduce the number of the arguments, we normalize the
partition function by the principal edge Boltzmann factors $x_0(K)$ and $%
x_0^*(K)$. 
\begin{equation}
\left\{ x_0(K) \right\}^{N_B} z(u_1(K)) = \left\{ x^*_0(K) \right\}^{N_B}
z(u^*_1(K)),
\end{equation}
where $z$ is the normalized partition function $z(u_1)=Z/\{x_0(K)\}^{N_B}$
and $z(u^*_1)=Z/\{x^*_0(K)\}^{N_B}$. We explicitly obtain $u_1(K) =
x_1(K)/x_0(K) = \exp(-2K)$ and $u_1^*(K) = x^*_1(K)/x^*_0(K) = \tanh K$. It
reads, if we set the dual coupling as $\exp(-2K^*) = \tanh(K)$, 
\begin{equation}
\left\{ x_0(K) \right\}^{N_B} z(u_1(K)) = \left\{ x^*_0(K) \right\}^{N_B}
z(u_1(K^*)).
\end{equation}
Therefore we find that the partition function for the Ising model has
symmetry between two different temperatures. This is the duality. The coefficient 
$\Lambda$ is also obtained as 
\begin{equation}
\Lambda = \left( \frac{x^*_0(K)}{x_0(K)} \right)^{N_B} = \left\{ \frac{1}{%
\sqrt{2}}\left(1 + \exp(-2K) \right) \right\}^{N_B}.
\end{equation}
The well known duality relation $\exp(-2K^*)=\tanh K$ can identify the exact
location of the critical point by equating $K=K^*$. The critical point is
estimated as $K_c = \ln(1 + \sqrt{2})/2 = 0.440686794\cdots$. Interestingly,
the coefficient $\Lambda$ becomes unity at the critical point. We use this
property as \textit{a priori} assumption for the analysis on the critical
point for spin glasses.

\subsubsection{Duality analysis: spin glass}

The replica method, which is often used in theoretical studies on spin
glasses, allows to generalize the duality analysis to spin glasses \cite%
{NN,MNN}. Let us consider the duality for the replicated partition function
as $[Z^n(K;\{\tau_{ij}\}]$. In this case, the multiple binary Fourier
transformation defines the dual edge Boltzmann factor as 
\begin{equation}
x^*_{k_1,k_2,\cdots,k_n} (K_p,K) = \left(\frac{1}{\sqrt{2}}%
\right)^n\sum_{\phi_1,\phi_2,\cdots,\phi_n}x_{\phi_1,\phi_2,\cdots,%
\phi_n}(K_p,K) \mathrm{e}^{\mathrm{i}\sum_{i=1}^{n}k_i \phi_i},
\end{equation}
where the original edge Boltzmann factor is given by the configurational
average of the $n$-replicated $\pm J$ Ising model, 
\begin{equation}
x_{\phi_1,\phi_2,\cdots,\phi_n} (K_p,K)= \frac{1}{2\cosh K_p}\left\{ \mathrm{%
e}^{-K_p + K \sum_{i=1}^n\cos \pi \phi_i } + \mathrm{e}^{K_p - K
\sum_{i=1}^n\cos \pi \phi_i }\right\}.
\end{equation}
Notice that the definition of $p$ is the density of the antiferromagnetic
interactions here to analyze the accuracy threshold for the surface code.
They leads us to the double expression of the replicated partition function
as, in a similar way to the above simple case, 
\begin{eqnarray}
& & \{x_0(K_p,K)\}^{N_B}z(u_1(K_p,K),u_2(K_p,K),\cdots)  \nonumber \\
& & = \{x^*_0(K_p,K)\}^{N_B}z(u^*_1(K_p,K),u^*_2(K_p,K),\cdots),
\end{eqnarray}
where the subscript of $u_k$ and $u_k^*$ stands for the number of
anti-parallel pair among $n$ replicas on each edge. We restrict ourselves to the case to analyze the location of the multicritical point, namely $K=K_p$
on the Nishimori line.

Unfortunately we cannot replace $u^*_k(K,K)$ by $u_k(K^*,K^*)$ even after normalization as the above
case for the Ising model, since the replicated partition function is
multivariable. Nevertheless we can estimate the location of the
multicritical point even without the ordinary procedure of the duality. We
here put a priori assumption that $x_0(K,K) = x^*_0(K,K)$ at the critical
point, implying the coefficient of the double expressions of the replicated
partition function should be unity. According to this assumption, we take
the limit $n \to 0$ of the equation along the replica method and thus obtain 
\begin{equation}
-p \log p -(1-p) \log (1-p) = \frac{1}{2}\log 2.
\end{equation}
The solution is $p_c = 0.1100\cdots$. We thus conclude that the accuracy
threshold is estimated as $p_c = 0.1100\cdots$. 
Strictly speaking, this analysis is not exact. In practical, the precision of the above result has been
confirmed to be satisfiable by supports from numerical simulations. In addition to such numerical validations, the following theoretical refinement of the method to
identify the multicritical point have been considered.

\subsubsection{Duality analysis with real-space renormalization}

\begin{figure}[tbp]
\begin{center}
\includegraphics[width=80mm]{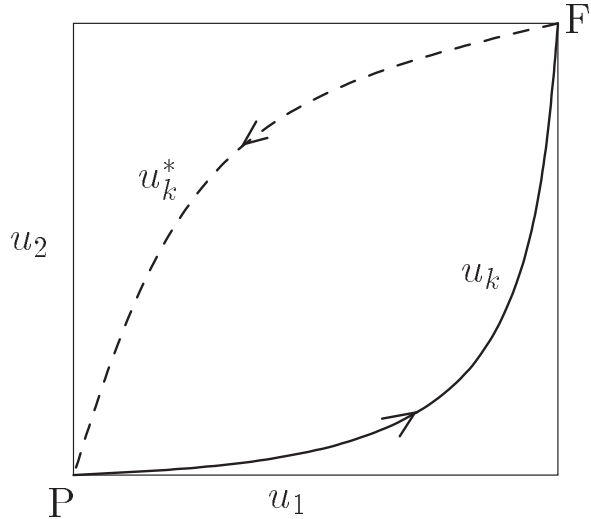}
\end{center}
\caption{Relative Boltzmann factors of the replicated partition function
(Projection onto two dimensions). The high-temperature limit is given by P,
while the low-temperature limit is $F$. The solid curve represents the
original relative Boltzmann factor. The dashed curve denotes the dual
relative Boltzmann factor. }
\label{FL}
\end{figure}

As above mentioned, the ordinary duality analysis hampers since the
replicated partition function was multivariable. Thus we rely on the
assumption that $x_0(K,K)=x^*_0(K,K)$ would give the location of the
critical point. We here sketch the relationship between the double
expressions of the replicated partition function as the curves of the
relative Boltzmann factors $u_k(K,K)$ and $u^*_k(K,K)$ on the
two-dimensional space for simplicity, although those are correctly in a
hyper space as in Fig. \ref{FL}. 
The thick curve denotes the relative Boltzmann factor $u_k(K,K)$ and the
dashed one represents the dual one $u^*_k(K,K)$. When the temperature
increases, the representative point of the replicated partition function
moves from P (the high-temperature limit) to F (the low temperature limit)
along the thick curve. On the other hand, the dual representative point goes
inversely along the dashed curve. These features have been shown rigorously
and imply the existence of the duality relation for the temperature \cite{ND}
. If two curves become completely coincident with each other, we then obtain
a relation implying $u^*_k(K,K)=u_k(K^*,K^*)$. Solving this relation, we can
obtain the duality relation for different temperatures as the case of the
Ising model. In spin glasses, they do not overlap.

\begin{figure}[tbp]
\begin{center}
\includegraphics[width=80mm]{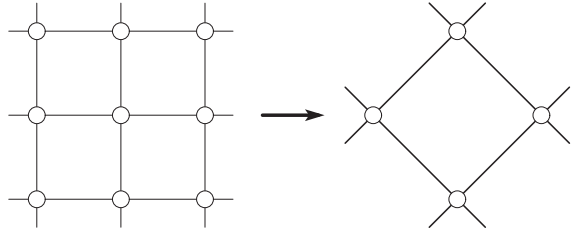}
\end{center}
\caption{ Real-space renormalization. In this instance, a limited number of
spins are summed, and the remaining ones construct another square lattice
with multiple-body interactions. }
\label{RS}
\end{figure}

\begin{figure}[tbp]
\begin{center}
\includegraphics[width=80mm]{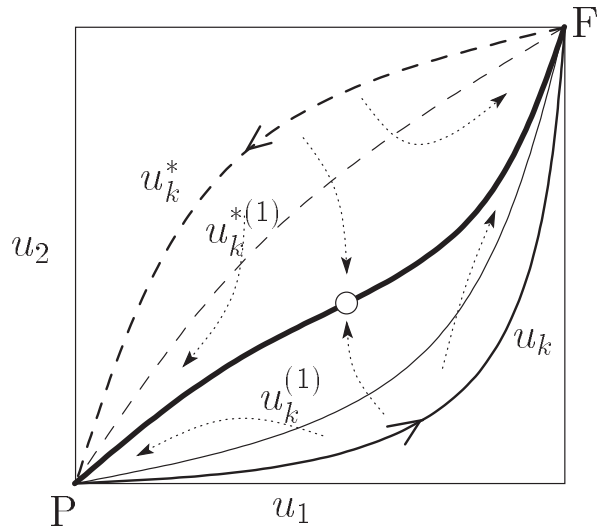}
\end{center}
\caption{Relative Boltzmann factors of the replicated partition function and
their renormalization flow (Projection onto two dimensions). The dotted
arrows depict the renormalization flow starting from several points on both
of the relative Boltzmann factors. The internal curves denote the
renormalized relative Boltzmann factors. The bold curve describes the
sufficiently renormalized relative Boltzmann factors with the unstable fixed
point as the case with a single variable in the partition function. }
\label{RFL}
\end{figure}

\begin{figure}[tbp]
\begin{center}
\includegraphics[width=100mm]{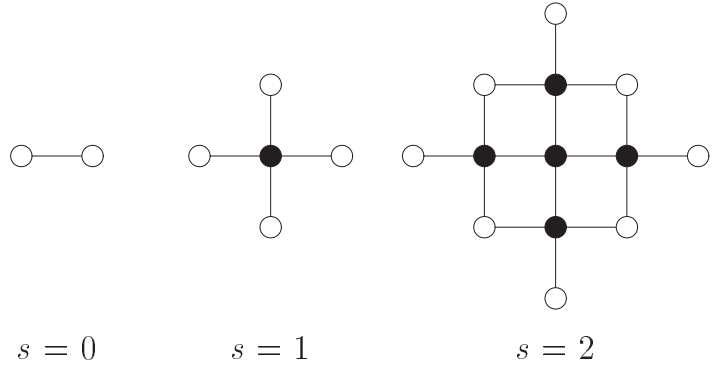}
\end{center}
\caption{Clusters for duality with renormalization. The white circles denote edge spins whose configuration is represented by $k$, and the black ones are to be summed.}
\label{DC}
\end{figure}

We import another piece of the theories in statistical mechanics, the
real-space renormalization group analysis. Most of the problems on
statistical mechanics are tractable since the degrees of freedom are highly
correlated. Often we employ a trick to map the original problem into much
simpler problem with a recursive structure by use of some approximation. As
in Fig. \ref{RS}, we trace over a part of spins on the square lattice. 
Repeating this procedure while omitting the generated multi-body
interactions (approximation), we construct the renormalization group in the
form of the recursion 
\begin{equation}
K_n = R(K_{n-1}),
\end{equation}
where $K_n$ is the renormalized coupling constant after $n$ steps. By use of
the renormalization group analysis, we describe flow of the renormalized
coupling constant in $K$ space. The flow usually terminates two fixed points
representing the ordered and disordered phase, while being divided by an
unstable fixed point, which is the critical point. The precision to
represent the original behavior in the model is dependent on the
approximation in construction of the renormalization group. We generalize
this procedure into the relative Boltzmann factors $\{u_k(K,K)\}$ and $
\{u^*_k(K,K)\}$. The flow can be depicted in the hyper space of $\{u_k(K,K)\}
$ as in Fig. \ref{RFL}. Similarly to the above simple case, the
renormalization flow goes toward two fixed points P and F. 
The unstable fixed point $C$ would be located between the original and dual
relative Boltzmann factors, since they do not overlap but have a common
critical point. The renormalization flow starts from both of the relative
Boltzmann factors and once moves to the unfixed point. After several
renormalization steps, the flow goes to two fixed points. That means that,
if we conduct one step of the renormalization for each point on the original
and dual relative Boltzmann factors, we observe both of the renormalized
curves get close to each other as in Fig. \ref{RFL}. The sufficient steps of
the renormalization makes both of the curves into a common thick curve
capturing the unfixed point, namely the multicritical point. As far as
possible, we desire to estimate the location of the critical point
precisely. Then, let us consider to trace the partial spins without any
approximations. For example, in the case on the square lattice, we define
the cluster Boltzmann factor $x_k^{(s)}(K,K)$, where the subscript $k$
denotes the configuration of the edge (white-colored) spins and $s$
expresses the size of the cluster in Fig. \ref{DC}. 
When $k=0$, all the edge spins are in up directions.
We sum over the internal (black-colored) spins in order to perform the
renormalization without approximation. It is difficult to perform the
sufficient step of the renormalization without any approximation. Remember
that we can estimate the relatively correct value of the multicritical point
even without the renormalization. Therefore we propose a systematic way to
improve precision of the location of the multicritical point. We employ the
following equation to estimate the location of the multicritical point, 
\begin{equation}
x_0^{(s)}(K,K) = x_0^{*(s)}(K,K).  \label{MCP}
\end{equation}
The equality for $s=0$ (edge) reproduces the case without renormalization as 
$p^{(0)}_c = 0.1100$ \cite{NN,MNN}. If we increase the size of the used
cluster, we can systematically approach the exact solution for the location
of the multicritical point of the $\pm J$ Ising model as $p^{(1)}_c = 0.1093$
and $p^{(2)}_c = 0.1092$.

If we remove the condition of the Nishimori line $K_p = K$, we can describe
the phase boundary for spin glasses by 
\begin{equation}
x_0^{(s)}(K_p,K) = x_0^{*(s)}(K_p,K).  \label{PhaseB}
\end{equation}
By use of this equation, we can obtain the precise results for the phase
boundary in the higher temperature region than the Nishimori line and the
whole phase boundary of the diluted Ising model \cite{ONB,Ohzeki}.

\subsubsection{Other cases}

Not only the surface code, several quantum error correcting codes are found
to possess the connection with the spin-glass models. Although we omit their
detailed explanations, we look over the recent results in short below. We
restrict ourselves to the case in which the duality analysis have
predicted the accuracy threshold.

\begin{figure}[tbp]
\begin{center}
\includegraphics[width=100mm]{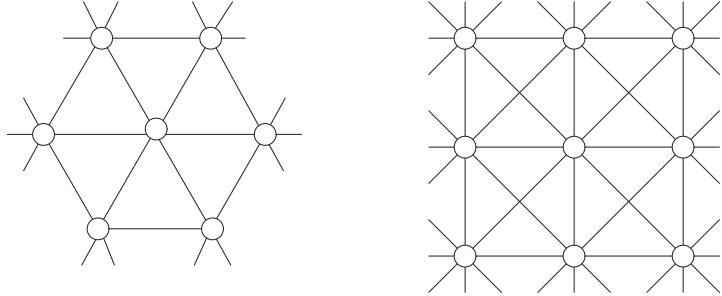}
\end{center}
\caption{Color code on the triangular and Union-Jack lattices.}
\label{CC}
\end{figure}

First, we simply mention the cases by other arrangement of physical qubits
in the surface code than the square lattice, say triangular and hexagonal
lattices. 
The accuracy threshold for both of the lattices can be given as $p_c = 0.164\cdots$ (triangular) and $p_c = 0.067\cdots$(hexagonal) \cite{Ohzeki}. 
Recently, another type of the surface code with more computational capability have been developed, color code. 
Also in the case of the color code, we prepare the arrangement of the physical qubits on each unit triangle on the triangular lattice or Union-Jack lattice as in Fig. \ref{CC}%
. \cite{Colorcode,Colorcode2}. 
The color code on the Union-Jack lattice implements the whole Clifford group of unitary gates generated by the Hadamard gate, the $\{\pi/8\}$ gate, and the controlled-NOT gate, although that on the triangular lattice can not employ the $\{\pi/8\}$ gate. 
It means that both of the color codes have the wider computational capability than that
of the Pauli group. 
The corresponding statistical mechanical model has three-body interactions differently from the above surface code. 
The quenched random interaction then represents the error arising on the physical qubits on the unit triangles. 
Here the duality analysis can estimate the accuracy thresholds for both of the color codes as $p_c=0.1096-8$ (triangular) and $p_c= 0.1092-3$ \cite{Ohzeki2}. Interestingly, the advantage of the computational capability does not spoil the robustness of the error correction.

\begin{figure}[tbp]
\begin{center}
\includegraphics[width=80mm]{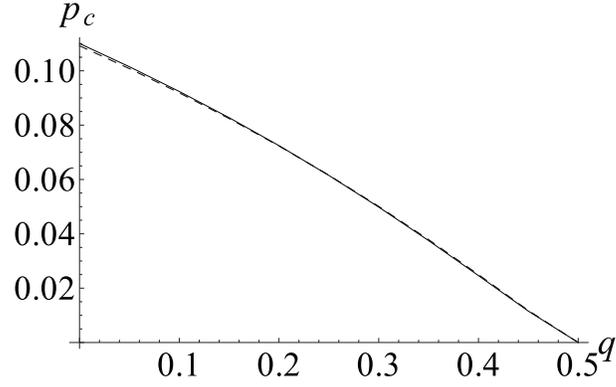}
\end{center}
\caption{Decay of the robustness due to loss of the physical qubits. 
The solid curve expresses the results by the duality with $s=0$ and the dashed
curve denotes those with $s=1$, while both of them are mostly overlapping in
this scale.}
\label{LQ}
\end{figure}

In addition, the practical errors in implementation of the quantum system are not only on the computational basis. 
Lack of the physical qubits is also likely to occur. 
Recently, Stace \textit{et al.} have proposed a modified scheme of the surface code to protect the logical quantum state from both type of the errors, namely from decoherence and lack of the physical qubits \cite{Loss1,Loss2}. 
Then the corresponding statistical mechanical model becomes the diluted version of the $\pm J$ Ising model. 
We can identify the locations of the optimal thresholds depending on the ratio of the loss of the physical qubits $q$ as in Fig. \ref{LQ} \cite{Ohzeki3}. 
The detailed values are shown in Table \ref{table1}. 
As shown in Table \ref{table1} and Fig. \ref{LQ}, the decay of the robustness, depending on increase of $q$, of the surface code against the flip/phase errors can be observed. 
This behavior can be interpreted as the decay of the ferromagnetic order due to
the dilution of the interactions. 
\begin{table}[htbp]
\tbl{Optimal thresholds given by the duality analyses with $s=0$, $1$, and $2$ clusters for the uncorrelated case.}
{
\begin{tabular}{lcccc}
\hline
$q$ & $p_c$ ($s=0$) & $p_c$ ($s=1$) & $p_c$ ($s=2$) &  \\ \hline
$0.00$ & $0.11003$ & $0.10928$ & $0.10918$ &  \\ 
$0.10$ & $0.09240$ & $0.09196$ & $0.09189$ &  \\ 
$0.20$ & $0.07245$ & $0.07235$ & $0.07233$ &  \\ 
$0.30$ & $0.04984$ & $0.05004$ & $0.05009$ &  \\ 
$0.40$ & $0.02462$ & $0.02492$ & $0.02500$ &  \\ 
$0.45$ & $0.01155$ & $0.01174$ & $0.01179$ &  \\ \hline
\end{tabular}%
}
\label{table1}
\end{table}

\subsubsection{Depolarizing channel}
Before closing this section, we add another example on the quantum error
correction related with statistical mechanics. 
We discussed the case without correlation of both of the flip and phase errors for simplicity. 
It is straightforward to generalize the above procedure for the depolarizing
channel case such that the error probabilities are equalized as $p_X=p_Y=p_Z=p/3$. In this case, the corresponding statistical mechanical model becomes a little bit different one from the $\pm J$ Ising model. 
We can no longer independently deal with both of the errors. 
Let us first evaluate the probability for the specific error chains. 
The error chains are generated following the distribution function as 
\begin{eqnarray}
P(E,E^*) &=& \prod_{\langle ij \rangle }\left(\frac{p}{3}\right)^{\frac{%
1+\tau^E_{ij}}{4}+\frac{1+\tau^{E^*}_{ij}}{4}+\frac{1+\tau^E_{ij}%
\tau^{E^*}_{ij}}{4}}(1-p)^{\frac{1-\tau^E_{ij}}{4}+\frac{1-\tau^{E^*}_{ij}}{4%
}+\frac{1-\tau^E_{ij}\tau^{E^*}_{ij}}{4}}  \nonumber \\
&\propto& \left(\frac{p}{3(1-p)}\right)^{\frac{\tau^E_{ij}}{4}+\frac{%
\tau^{E^*}_{ij}}{4}+\frac{\tau^E_{ij}\tau^{E^*}_{ij}}{4}}.
\end{eqnarray}
We again use the expression of the exponential form as in the case for the
spin glass and obtain 
\begin{equation}
P(E,E^*) \propto \prod_{\langle ij
\rangle}\exp(K_d\tau^E_{ij}+K_d\tau^{E^*}_{ij}+K_d\tau^E_{ij}%
\tau^{E^*}_{ij}).
\end{equation}
Notice that $\exp(-4K_d) = 3(1-p)/p$. 
Similarly, the problem is to identify the singularity of the following the probability through the partition function of a spin glass 
\begin{eqnarray}
& &P_K(\partial E, \partial E^* | \bar{E},\bar{E}^*)  \nonumber \\
& &\quad \propto \sum_{_{S_i,S_i^*}}\prod_{\langle
ij\rangle}\exp(K\tau^{E}_{ij}S_{i}S_{j}+K\tau^{E^*}_{ij}S^*_iS^*_j+K%
\tau^{E}_{ij}\tau^{E^*}_{ij}S_iS_jS^*_iS^*_j),  \nonumber \\
\end{eqnarray}
where the quenched random coupling obeys the distribution function $P(E,E^*)$
. 
This model is regarded as a spin-glass version of the $4$-state Potts model.

The duality analysis can be performed for this statistical-mechanical model
in order to estimate the optimal thresholds \cite{AHMMH}. 
In addition, we can apply the modified scheme to circumvent the effects due to loss of the physical qubits proposed by Stace \textit{et al.} \cite{Ohzeki3}. 
We list the results for the depolarizing channel in Table \ref{table2}.
\begin{table}[htbp]
\tbl{Results for the correlated case (depolarizing channel) $
p_X=p_Z=p_Y=p/3$.}
{\begin{tabular}{lccc}
\hline
$q$ & $p_c$ ($s=0$) & $p_c$ ($s=1$) &  \\ \hline
$0.00$ & $0.18929$ & $0.18886$ &  \\ 
$0.10$ & $0.16025$ & $0.15985$ &  \\ 
$0.20$ & $0.12690$ & $0.12656$ &  \\ 
$0.30$ & $0.08844$ & $0.08819$ &  \\ 
$0.40$ & $0.04454$ & $0.04440$ &  \\ 
$0.45$ & $0.02121$ & $0.02114$ &  \\ \hline
\end{tabular}}
\label{table2}
\end{table}

\section{Quantum annealing and beyond}

The second part of this chapter, we introduce another application of
statistical mechanics in context of quantum computation. 
As in the other chapter, by use of the analogy of simulated annealing (SA) developed in
statistical mechanics, a generic algorithm intended for solving optimization problems by use of quantum nature, quantum annealing (QA), has been studies since its proposal. In SA, we make use of thermal, classical, fluctuations to employ a stochastic search for the desired lowest-energy state, which corresponds to the optimal solution, by allowing the system to hop from state to state over intermediate energy barriers. 
In QA, by contrast, we introduce non-commutative operators as artificial degrees of freedom of quantum nature in order to induce quantum fluctuations. 
The most typical procedure of QA is performed by adiabatic control of quantum fluctuations, say quantum adiabatic computation (QAC). 
The adiabatic theorem guarantees that an sufficient slow time evolution would let the system closely follow the instantaneous ground state. 
Therefore, starting from an initial trivial ground state, we can reach a nontrivial one by slowly controlling quantum fluctuations. 
However, in a practical sense, we desired to perform the procedure to find the answer of the optimization problem as fast as possible. The adiabatic control to efficiently obtain the optimal solution takes a characteristic time related with the energy gap between the first excited state and the lowest one during its procedure. 
The problems with closure of the energy gap are difficult to be solved in a moderate time by QAC. 
Unfortunately, as far as we know, the typical difficult problems in classical computation involves the exponential closure of the energy gap by increase of the problem size.

In this chapter, we show several trials to overcome the bottleneck of quantum adiabatic computation by use of statistical mechanics. 
We use formal similarities between SA and QAC in order to import several developments in statistical mechanics. In both methods, we have to drive the system slowly and carefully to control the strengths of thermal or quantum fluctuations. 
The idea behind quantum adiabatic computation is to keep the system close to the instantaneous ground state of a quantum system. 
This is analogous to the protocol of SA, in which one tries to make the system keep quasi-equilibrium state. It is indeed possible to make this analogy more transparent by a precise formulation, from which a fascinating technique to be expected to improve the performance of QAC can be proposed.

\subsection{Quantum adiabatic computation: short review}

Let us briefly review the procedure of the most typical QA, namely QAC,
before importing theory of statistical mechanics to quantum computation. 
In QA, we introduce a non-commutative operator to drive the system by quantum nature as 
\begin{equation}
H(t) = f(t)H_0 + \left(1 - f(t)\right) H_1,  \label{QAH}
\end{equation}
where $H_0$ is the classical Hamiltonian consisting of diagonal elements, which express the cost function. 
Here $f(t)$ is assumed to be a monotonically increasing function satisfying $f(0) = 0$ and $f(T) = 1$. 
For instance, $f(t) = t/T$, where $T$ denotes the computation time for QAC.
 The quantum annealing starts from a trivial ground state of $H_1$, which is often chosen to be a uniform linear combination of the computational basis as $|\Psi(0)\rangle = |\sigma \rangle /\sqrt{N}$. 
Let us deal with only the discrete combinatorial optimization problems, which are simply termed as the optimization problem below. 
Most of the optimization problem can be expressed by the spin-glass Hamiltonian. 
In order to explain the procedure of QAC in detail, we consider to find the ground state of the simple spin-glass model as used before, namely the random-bond Ising model, 
\begin{equation}
H_0 = - \sum_{\langle ij \rangle} J_{ij} \sigma_i^z \sigma_j^z,
\end{equation}
where the summation is taken over all the nearest-neighboring pairs of the Ising spins. We take the computational basis of the eigenstates of the Ising variables to represent the instantaneous state as $|\Psi(t) \rangle =|\sigma_1^z,\sigma_2^z,\cdots,\sigma_N^z \rangle$. 
The transverse-field operator is often used as quantum fluctuations for implementing QAC for the spin-glass model 
\begin{equation}
H_1 = - \Gamma _{0} \sum_{i=1}^N \sigma_i^x,
\end{equation}
where $\Gamma _{0}$ is the strength of the transverse field. 
The whole Hamiltonian of QAC (although widely used for QA) thus becomes 
\begin{equation}
H(t) = f(t)\sum_{\langle ij \rangle}J_{ij}\sigma^z_i\sigma_j^z + \left(1 -
f(t)\right)\Gamma_0 \sum_{i=1}^N \sigma^x_i.  \label{QAS}
\end{equation}
The ground state of the transverse-field operator $H_1$ is trivially given by a uniform linear combination as $\sum_{\{\sigma\}} | \sigma \rangle/\sqrt{2}^N$. 
For a sufficiently large $T$, the adiabatic theorem guarantees that the instantaneous state at time $t$, $|\Psi(t)\rangle$, is very close to the instantaneous ground state implying $|0(t)\rangle$, $\langle 0(t)|\Psi(t)\rangle \approx 1$, when the instantaneous ground state $|0(t)\rangle$ is non-degenerate. 
The condition for $|0(t)\rangle$, $\langle 0(t)|\Psi(t)\rangle 1 - \epsilon^2 (\epsilon \ll 1)$ to hold is given by 
\begin{equation}
\frac{\mathrm{max}\left|\langle 1(t)| \frac{dH(t)}{dt}|0(t)\rangle\right|}{\mathrm{min}\Delta^2(t)} = \epsilon,
\end{equation}
where $|1(t)\rangle$ is the instantaneous first excited state, and $\Delta(t)
$ is the energy gap between the ground state and first excited one. The
maximum and minimum are evaluated between $0$ and $T$. In our case, since $%
dH(t)/dt \propto 1/T$, the adiabatic condition is written as 
\begin{equation}
T \propto \frac{1}{\epsilon \mathrm{min}\Delta^2(t)}.
\end{equation}
Therefore, if we desire to solve the problems involved with the exponential
closure of the energy gap while increase of $N$, QAC must take extremely
long time to find the ground state with high probability \cite{FT1,FT2}. 

\subsection{Novel type of quantum annealing}
In order to overcome this problematic bottleneck of QAC, we change our strategy
from the adiabatic control of quantum fluctuations. We demand an important
key for nonequilibrium statistical mechanics by using a fascinating bridge
between quantum computation as QAC and statistical mechanics. 
To make the connection more clear, we show a useful technique to relate both of the
fields.

\subsubsection{Classical quantum mapping}

Let us compare two of the procedures of SA and QAC. Both of the protocols
are given by slow sweep of thermal and quantum fluctuations to keep the
system trace the instantaneous stationary state, equilibrium for SA and
ground state for QAC, respectively. In numerical implementation of SA, we
demonstrate a stochastic dynamics driven by thermal fluctuation with the
master equation. 
\begin{equation}
\frac{d}{dt}P(\sigma;t) = \sum_{\sigma^{\prime }} M(\sigma|\sigma^{\prime
};t)P(\sigma^{\prime };t),
\end{equation}
where $P(\sigma;t)$ is the probability with a spin configuration of $%
\{\sigma_i^z\}$ simply denoted as $\sigma$ at time $t$. Notice that $\sigma$
is not a state of a single spin but is a collection of spin states. $%
M(\sigma^{\prime }|\sigma;t)$ expresses the transition matrix following the
conservation of probability $\sum_{\sigma} M(\sigma|\sigma^{\prime };t) = 1 $
and the detailed balance condition 
\begin{equation}
M(\sigma|\sigma^{\prime };t)P_{\mathrm{eq}}(\sigma^{\prime };t) =
M(\sigma^{\prime }|\sigma;t)P_{\mathrm{eq}}(\sigma;t),
\end{equation}
where we denote the instantaneous equilibrium distribution as $P_{\mathrm{eq}%
}(\sigma;t) = \exp(-\beta(t)E(\sigma;t))/Z(t)$ and the instantaneous energy $%
E(\sigma;t)$ is the value of the classical Hamiltonian $H_0(t)$. Since SA
consists of a dynamic control of the temperature, the time variable $t$ has
been written explicitly in the arguments of the inverse temperature and the
partition function. In order to satisfy this condition, we often use the
transition matrix with Metropolis rule as 
\begin{equation}
M(\sigma|\sigma^{\prime };t) = \mathrm{min}(1, \exp(-\beta \Delta E(\sigma|
\sigma^{\prime };t))),
\end{equation}
where 
\begin{equation}
\Delta E(\sigma| \sigma^{\prime };t) = E(\sigma;t) - E(\sigma^{\prime };t),
\end{equation}
or heat-bath rule as 
\begin{equation}
M(\sigma|\sigma^{\prime };t) =\delta_1(\sigma,\sigma^{\prime }) \frac{%
\exp\left(-\frac{\beta}{2}\Delta E(\sigma|\sigma^{\prime };t) \right)}{%
2\cosh \left(\frac{\beta}{2}\Delta E(\sigma|\sigma^{\prime };t)\right)},
\end{equation}
where 
\begin{equation}
\delta_1(\sigma|\sigma^{\prime }) = \delta(2,\sum_{i=1}^N(1-\sigma_i
\sigma^{\prime }_i)).
\end{equation}

On the other hand, the dynamics of QAC is governed by the Shrodinger
equation. To look at the connection between SA and QAC, we employ a mapping
technique of dynamics of relaxation toward equilibrium as the master
equation into the Shrodinger equation. If we use the following special
quantum Hamiltonian, we find it possible to simulate the dynamics of SA in
quantum manner \cite{QC}, 
\begin{equation}
H_{q}(\sigma ^{\prime }|\sigma ;t)=\delta (\sigma ^{\prime },\sigma )-%
\mathrm{e}^{\beta (t)H_{0}(\sigma ^{\prime })/2}M(\sigma ^{\prime }|\sigma
;t)\mathrm{e}^{-\beta (t)H_{0}(\sigma )/2}.
\end{equation}%
Here we consider to gradually increase the inverse temperature following the
spirit of SA as $\beta (t)$. This special Hamiltonian has the following
state as its ground state, 
\begin{equation}
|\Psi _{\mathrm{eq}}(t)\rangle =\frac{1}{\sqrt{Z(t)}}\sum_{\sigma }\mathrm{e}%
^{-\frac{\beta (t)}{2}H_{0}(\sigma )}|\sigma \rangle .
\end{equation}%
It is easy to confirm that the quantum expectation value of a physical
quantity $A(\sigma )$ in the ground state as $\langle \Psi _{\mathrm{eq}%
}(t)|A(\sigma )|\Psi _{\mathrm{eq}}(t)\rangle $ coincides with the thermal
expectation of the same quantity with $\beta (t)$. The ground state energy
simply takes zero. This fact is shown by use of the conservation of the
probability and the detailed-balance condition as 
\begin{eqnarray}
&&\left( \delta (\sigma ^{\prime },\sigma )-\mathrm{e}^{\frac{\beta (t)}{2}%
H_{0}(\sigma ^{\prime })}M(\sigma ^{\prime }|\sigma ;t)\mathrm{e}^{-\frac{%
\beta (t)}{2}H_{0}(\sigma )}\right) |\Psi _{\mathrm{eq}}(t)\rangle  
\nonumber \\
&\propto &\sum_{\sigma }\left( \mathrm{e}^{\frac{\beta (t)}{2}H_{0}(\sigma
^{\prime })}-\mathrm{e}^{\frac{\beta (t)}{2}H_{0}(\sigma ^{\prime
})}M(\sigma ^{\prime }|\sigma ;t)\right) |\sigma \rangle =0.
\end{eqnarray}%
On the other hand, the excited states have positive-definite eigenvalues,
which can be confirmed by the application of the Perron-Frobenius theorem.
By using the above special quantum system, we can treat a quasi-equilibrium
stochastic process in SA as an adiabatic dynamics as in QAC.

The above formulation is a generic way of the classical-quantum mapping. We
demonstrate the above mapping of SA into QAC by more explicit instance. Let
us consider an optimization problem that can be expressed as a classical
Hamiltonian with local interaction 
\begin{equation}
H_0 = - \sum_{j} H_j,  \label{SAc}
\end{equation}
where $H_j$ involves $\sigma_j^z$ and a finite number of $\sigma_k^z (k _neq
j)$. Taking a familiar instance is a spin-glass system with nearest-neighbor
interactions, 
\begin{equation}
H_j = - \sum_{k \in \partial j} J_{jk} \sigma_j^z\sigma_k^z,
\end{equation}
where $\partial j$ denote sites adjacent to $j$. The following Hamiltonian
is the explicit form, which facilitates our analysis, 
\begin{equation}
H_{\mathrm{q}}^{\mathrm{SG}}(t) = - \chi(t) \sum_{j}\left( \sigma^x_j - 
\mathrm{e}^{\frac{\beta(t)}{2}H_j}\right),  \label{SAq}
\end{equation}
where $\chi(t) = \mathrm{e}^{\beta(t)p}$ with $p = \mathrm{max}_i|H_i |$.
Notice that $p$ is proportional to the interaction energy and is the order
of $\mathcal{O}(N^0)$ due to finiteness of the interaction range. Let us
consider that the protocol of SA in the above quantum system. In the first
stage with very high temperature $\beta(0) \to 0$, the quantum system (\ref%
{SAq}) reduces 
\begin{equation}
H_{q}^{\mathrm{SG}}(t) = - \sum_j \left( \sigma_j^x - 1 \right).
\end{equation}
Its ground state is the uniform linear combination of all possible states in
the basis to diagonalize $\{\sigma_i^z\}$. It means that all states appear
with an equal probability as the equilibrium distribution in
high-temperature limit. Therefore the quantum system (\ref{SAq}) can
correctly demonstrate the initial condition of SA. In addition, in the limit 
$\beta(t \to T) \to \infty$, (\ref{SAq}) becomes 
\begin{equation}
H_{q}^{\mathrm{SG}}(t) \sim \chi(t) \sum_j \mathrm{e}^{\frac{\beta(t)}{2}%
H_j}.
\end{equation}
The ground state is with the lowest value of the classical system (\ref{SAc}%
), because each $H_j$ takes its lowest value. These observations confirm
that the quantum system (\ref{SAq}) indeed demonstrate the quasi-stationary
dynamics in SA. Notice that, Interestingly, the adiabatic condition of the
above special quantum system we used in the classical-quantum mapping can
reproduce the condition of convergence of SA. By use of the fascinating
connection between SA and QAC, we can import several theories of statistical
mechanics into the quantum dynamics.

As shown later, the collaboration of statistical mechanics with quantum
dynamics can produce a new algorithm to solve optimization problems in
different manners from QAC.

\subsubsection{Jarzynski equality}

Among several recent developments in statistical mechanics, we take the
Jarzynski equality (JE) as an attempt to improve the performance of QAC \cite%
{PRL}. This equality relates quantities at two different thermal equilibrium
states with those of nonequilibrium processes connecting these two states.
It can also be termed as a generalization of the well-known inequality, the
second law of thermodynamics $\Delta F \le \langle W \rangle_{0 \to T}$.
Here the brackets $\langle \cdots \rangle_{0 \to T}$ are for the average
taken over nonequilibrium processes between the initial (at $t=0$) and final
states (at $t=T$), which are specified only macroscopically and thus there
can be a number of microscopic realizations.

The Jarzynski equality is written as \cite{J1,J2} 
\begin{equation}
\left\langle \mathrm{e}^{-\beta W} \right\rangle_{0 \to T} = \frac{Z_{T}}{Z_0%
}.
\end{equation}
Here the partition functions for the initial and final Hamiltonians are
expressed as $Z_0$ and $Z_T$ , respectively. One of the important features
is that JE holds independently of the pre-determined schedule of the
nonequilibrium process. Another celebrated benefit is that JE reproduces the
second law of thermodynamics by using the Jensen inequality. Notice that we
have to take all fluctuations into account in evaluation of the expectation
value in the right-hand side of JE in order to calculate the free energy
difference. The Jarzynski equality holds formally in the case with change of
temperature, 
\begin{equation}
\left\langle \mathrm{e}^{-Y} \right\rangle_{0 \to T} = \frac{Z_{T}}{Z_0},
\end{equation}
when we employ the pseudo work instead of the ordinary performed work due to
the energy difference as 
\begin{equation}
Y(\sigma;t_k) = \left(\beta_{k+1} - \beta_k\right)E(\sigma;t_i).
\end{equation}
Here we employ discrete time expressions as $t_0 = 0$ and $t_n = T$ for
simplicity. We show the simple proof of JE for the dynamics in SA. Let us
consider a thermal nonequilibrium process in a finite-time schedule governed
by the master equation. The left-hand side of JE is explicitly written as 
\begin{equation}
\left\langle \mathrm{e}^{-Y} \right\rangle_{0 \to T} = \sum_{\{ \sigma_k \}}
\prod_{k=0}^{n-1} \left\{ \mathrm{e}^{- Y(\sigma_{k+1};t_k)} \mathrm{e}%
^{\delta t M(\sigma_{k+1}|\sigma_k;t_k)} \right\} P_{\mathrm{eq}%
}(\sigma_0;t_0).  \label{CJE}
\end{equation}
We take the first product of the above equation as, 
\begin{eqnarray}
&&\sum_{\sigma _{0}}\left\{ \mathrm{e}^{-Y(\sigma _{1};t_{0})}\mathrm{e }%
^{\delta tM_{0}(\sigma _{1}|\sigma _{0};t_{0})}\right\} P_{\mathrm{eq}%
}(\sigma _{0};t_{0})  \nonumber \\
&&\quad =P_{\mathrm{eq}}(\sigma _{1};t_{1})\frac{Z_{1}}{Z_{0}}.
\end{eqnarray}
Repetition of the above manipulation in Eq. (\ref{CJE}) yields the quantity
in the right-hand side of JE as, 
\begin{equation}
\sum_{\sigma _{n}}P_{\mathrm{eq}}(\sigma _{n};t_{n})\prod_{k=0}^{n-1}\frac{%
Z_{k+1}}{Z_{k}}=\frac{Z_{n}}{Z_{0}},
\end{equation}
where $Z_n=Z_T$. This is the case for a classical system on a heat bath, not
for a quantum system. Although readers may think the above proof is not
relevant to improvement of QAC, the formulation of JE for the classical
system is available for QAC by aid of the classical-quantum mapping above
introduced.

\subsubsection{Quantum Jarzynski annealing}

We here provide a novel protocol from the same spirit as JE by using the
special quantum system (\ref{SAq}). Let us consider to start from the
trivial ground state with the uniform linear combination similarly to the
case of the ordinary QA. This initial state expresses the high-temperature
equilibrium state as $|\Psi_{\mathrm{eq}} (t_0)\rangle \propto \mathrm{e}%
^{-\beta(t_{0})H_{0}(\sigma)/2}|\sigma \rangle $ with $\beta(t_0)\ll 1$. We
introduce the exponentiated pseudo work operator $R(t_k)=\exp(-Y(%
\sigma_k;t_k)/2)$. Observant readers might think it as a non-unitary
operator, but we can construct this operation by considering an enlarged
quantum system as detailed later. When we apply $R(t_k)$ to $|\Psi_{\mathrm{%
eq}} (t_k)\rangle$ with the inverse temperature $\beta(t_k)$, the quantum
state is changed into a state corresponding to the equilibrium distribution
with $\beta (t_{k+1})$. Then the application of the time-evolution operator $%
U(\sigma^{\prime}|\sigma;t_{k+1})=\exp (-\mathrm{i}\delta t
H_{q}(\sigma^{\prime}|\sigma;t_{k+1})/\hbar)$ does not alter the
instantaneous quantum state, since it is the ground state of $%
H_{q}(\sigma^{\prime}|\sigma;t_{k+1})$. The resultant state after the
repetition of the above procedure is 
\begin{eqnarray}
|\Psi (t_{n})\rangle & \propto & \prod_{k=1}^{n}\left\{
R(t_{k})U_{k}(\sigma_{k}|\sigma_{k-1};t_{k}) \right\}|\Psi_{\mathrm{eq}}
(t_{0})\rangle .  \nonumber \\
\end{eqnarray}
The product in the right-hand side is essentially of the same form as in Eq.
(\ref{CJE}). Instead of the exponentiated matrix of $\delta t
M(\sigma_{k+1}|\sigma_{k};t_{k})$, we use the time-evolution operator $%
U_k(\sigma_{k}|\sigma_{k-1};t_{k})$ here. After the system reaches the state 
$|\Psi (t_n)\rangle$, we measure the obtained state by the projection onto a
specified state $\sigma ^{\prime }$. The probability is then given by $%
|\langle \sigma^{ \prime }|\Psi(t_{n})\rangle |^{2}$, which means that the
desired ground state can be obtained with the probability proportional to $%
\exp(-\beta(t_n)H_{0})$, since $|\Psi (t_{n}) \rangle \propto |\Psi_{\mathrm{%
eq}} (t_{n})\rangle$. If the above procedure continues up to $\beta (t_n)\gg
1$, the resultant wave function can yield the ground state of $H_{0}$. We
call this procedure as the quantum Jarzynski annealing (QJA).

We here emphasize the following three points. First, the protocol of QJA
does not rely on the quantum adiabatic control. The computational time does
not depend on the energy gap. In this sense, QJA does not suffer from the
energy-gap closure differently from QAC. The required computational cost for
realization of QJA is estimated from the number of the unitary gates as will
be discussed below. Second, from the property of JE, the result is
independent of the schedule to tune the parameter, $T$, in the above
manipulations. Third, we do not need the repetition of the pre-determined
process to deal with all fluctuations in the nonequilibrium-process average
as in the ordinary JE, since the classical ensemble is mapped to the quantum
wave function. In addition, if we obtain the final wave function, the output
can give the ground state we desire with a very high probability since $%
\beta(t_n) \gg 1$. \footnote{%
Notice that several-time repetitions of experiments should be demanded since
the output by quantum measurement is probabilistic. However we should
emphasize that this point is not related with the theoretical property of JE
attributed to rare events, necessity of all the realizations during the
nonequilibrium process, but it comes from quantum nature.}

\subsubsection{Problems in measurement of answer}

So far, so good. No problems seem to exist in the realization and
performance of QJA. Unfortunately, we can find a serious problem to
efficiently solve the hard optimization problem by QJA. In order to
implement QJA, we must prepare a peculiar operator, the exponentiated pseudo
work operator $R(t_{k})=\exp (-Y(\sigma;t_k)/2)$, which looks like a
non-unitary operator. We can realize this non-unitary operator for the
original Hilbert space by preparing an enlarged quantum system with an
ancilla qubit (another two-level quantum system) as $|\Psi ,\phi_1 \rangle =
|\Psi \rangle \otimes |\phi_1 \rangle $, where $\phi_1$ is assumed to take $0
$ and $1$ \cite{WQ}. We call the ancila qubit the computational state below.
We initially set $|\Psi ,\phi_1=0 \rangle$. For simplicity, we restrict
ourselves to the case with $H_0(\sigma) > 0$ for any states. Let us
introduce the following unitary operator for the enlarged quantum system as 
\begin{eqnarray}
R^{\mathrm{unit.}}(t_k) &=& \sum_{\sigma }|\sigma \rangle \langle \sigma
|\otimes \left( 
\begin{array}{cc}
\sqrt{y_k(\sigma )} & \sqrt{1-y_k(\sigma )} \\ 
-\sqrt{1-y_k(\sigma )} & \sqrt{y_k(\sigma )}%
\end{array}%
\right)  \nonumber \\
&\equiv& I_{\sigma} \otimes Y_k,
\end{eqnarray}
where $y_k(\sigma )=\exp (-Y_k(\sigma;t_k))$. We can obtain the weighted
quantum system by applying this operator as $\sqrt{y_k(\sigma )}|\Psi
,\phi_1 =0\rangle $. Then we regard $R^{\mathrm{unit.}}(t_k)$ as the
exponentiated pseudo work operation $R(t_{k})$ for the quantum state $|\Psi
,\phi_1 =0\rangle $. The other probability amplitudes of $R^{\mathrm{unit.}%
}|\Psi ,\phi_1=0\rangle $ are given as 
\begin{eqnarray}
\langle \Psi ,0|R^{\mathrm{unit.}}(t_k)|\Psi ,0\rangle &=&\sqrt{y_k(\sigma )}
\\
\langle \Psi ,1|R^{\mathrm{unit.}}(t_k)|\Psi ,0\rangle &=&\sqrt{1-y_k(\sigma
)}.
\end{eqnarray}
The output in measurement of the quantum state, $\sqrt{1-y_k(\sigma )}%
|\Psi,\phi_1 =1\rangle $ is regarded as an undesired error state in our
computation. Therefore we here do not employ such error states to find the
ground state of the classical Hamiltonian $H_0(\sigma)$. Let us assume the
uniform change of the inverse temperature $\beta(t_{k+1}) - \beta(t_k)
\equiv \delta \beta $ for simplicity. In order to gain the relevant weight
for obtaining the desired ground state of $H_{0}$, we must increase a
parameter corresponding to the inverse temperature up to $%
\beta(t_{n})\epsilon \sim 1$, where $\epsilon $ is the minimum energy gap of
the ``classical" Hamiltonian $H_{0}$ (usually given by the energy unit).
Therefore the number of steps of QJA is necessary up to $n \equiv
\beta(t_{n})/ \delta \beta \sim 1/ \epsilon \delta \beta $. The
computational time $n$ is the size of the enlarged quantum state we have to
prepare as $|\Psi ,\phi _{1},\cdots ,\phi _{n}\rangle $. The initial state
of QJA in this case is $|\Psi_{\mathrm{eq}}(t_0) ,0,\cdots ,0\rangle $. The
other states as $|\Psi,1,0\cdots ,0\rangle $, $|\Psi ,0,1,\cdots ,0\rangle $%
, etc. such that several ancilla qubits take different states from the
initial ones as $\phi _{i}=1$ are regarded as the error states. To gain the
weight up to $\exp(-\beta (t_{n})H_{0})$, we perform the $n$-step
exponentiated work operations as $I_{\sigma }\otimes Y_{1}\otimes
I_{2}\otimes \cdots \otimes I_{n}$, $I_{\sigma }\otimes I_{1}\otimes
Y_{2}\otimes I_{3}\otimes \cdots \otimes I_{n}$, $\cdots $ and $I_{\sigma
}\otimes I_{1}\otimes \cdots \otimes I_{n-1}\otimes Y_{n}$, where $I_{j}$
denotes the identity operation. We can then obtain the desired state $|\Psi
,0, 0, \cdots, 0 \rangle $ with the weight as $\exp (-\beta (t_{n})H_{0})$.
The weights for the other states, the error states, are given as $%
(1-\exp(-\beta (t_{n})H_{0}))^{1/n}(\exp(-\beta (t_{n})H_{0}))^{1-1/n}$ for $%
|\Psi ,1,0,\cdots ,0\rangle $ and $(1-\exp(-\beta
(t_{n})H_{0}))^{2/n}(\exp(-\beta (t_{n})H_{0}))^{1-2/n}$ for $|\Psi
,1,1,0,\cdots ,0\rangle $ and so on.

Let us consider to obtain the meaningful outputs by the projective
measurements. The desired state is simply one with $\phi_i = 0$ for any $i$.
The probability for obtaining the state $|\sigma,0,0,\cdots, 0\rangle$ is
evaluated as 
\begin{equation}
p_0(\sigma) = \frac{\mathrm{e}^{-\beta (t_{n})H_{0}(\sigma)}}{%
\sum_{\sigma}\sum_{k=0}^n (1-\mathrm{e}^{-\beta (t_{n})H_{0}(\sigma)})^{k/n}(%
\mathrm{e}^{-\beta (t_{n})H_{0}(\sigma)})^{1-k/n}} = \frac{\mathrm{e }%
^{-\beta (t_{n})H_{0}(\sigma)}}{2^N}.
\end{equation}
Thus the successful result is generated with the exponentially decreasing
probability for increase of the system size $N$. We will demand more
ingenious techniques to efficiently obtain the desired outputs. This is the
remaining problem on QJA.

\subsection{Non-adiabatic quantum computation}

In the previous section, we consider to implement the manipulation in the
left-hand side of JE in the quantum computation by use of the
classical-quantum mapping. On the other hand, we have another direction of
QA to improve the performance beyond QAC by directly considering the
nonequilibrium behavior of quantum system. Again, in order to develop the
theory of QA, we rely on the property of JE but for the quantum system.
Instead of the adiabatic control, we consider to repeat non-adiabatic
quantum annealing (small or intermediate $T$) starting from a state chosen
from equilibrium ensemble, not necessarily the ground state. We may not be
able to easily reach the ground state of $H_0$ by such processes, since the
system does not keep the instantaneous ground state as in the adiabatic
computation. We instead need to repeat the process many times to hit the
ground state. In this way, the problem of long annealing time is expected to
be replaced by many repetitions of non-adiabatic (possibly quick) evolution.
We call such a procedure as non-adiabatic quantum annealing (NQA) \cite{NQA}.

\subsubsection{Jarzynski equality for quantum system}

Before analysis on the detailed property of the non-adiabatic computation,
we recall the Jarzynski equality but for isolated quantum system \cite{QJE1,QJE2}, while assuming its application to NQA.

Let us consider to find the ground state as of the spin-glass Hamiltonian $%
H_0$ as in Eq. (\ref{SAc}) by NQA. We prepare the dynamical quantum system
following the time-dependent Hamiltonian (\ref{QAH}). Initially we pick up a
state from the canonical ensemble for $H(0)=H_1=-\Gamma_0\sum_i\sigma_i^x$
and then let it evolve following the time-dependent Schr\"{o}dinger
equation. The performed work in the isolated quantum system is given by the
difference between the outputs of projective measurements of the initial and
final energies, $W=E_{m}(T)-E_{n}(0)$. Here $m$ and $n$ denote the indices
of the instantaneous eigenstates measured at the final and initial steps of
NQA, $H(T)|m(T)\rangle =E_{m}(T)|m(T)\rangle $ and $H(0)|n(0)%
\rangle=E_{n}(0)|n(0)\rangle $, respectively. The time-evolution operator is
given by the following unitary operator as 
\begin{equation}
U_{T} = \mathcal{T} \exp\left(\mathrm{i}\int_{0}^{T}dt H(t)\right),
\end{equation}
where $\mathcal{T}$ denotes the time ordered product. Thus we can evaluate
the transition probability between the initial and final steps as 
\begin{equation}
P_{m,n}(0 \to T) = |\langle \Psi_m(T)| U_T | \Psi_n(0) \rangle |^2.
\end{equation}
Therefore the path probability for the nonequilibrium process starting from
the equilibrium ensemble as 
\begin{equation}
P_{m,n}(0 \to T)\frac{\exp(-\beta E_n(0))}{Z_0(\beta; \{J_{ij}\})},
\end{equation}
where we express the instantaneous partition function at each time $t$ as $%
Z_t(\beta;\{J_{ij}\})$. By directly evaluating the left-hand side of JE, we
reach JE for the isolated quantum system as 
\begin{eqnarray}
\left\langle \mathrm{e}^{-\beta W} \right\rangle_{\mathrm{QA}} &=&
\sum_{m,n} \mathrm{e}^{-\beta W }P_{m,n}(0 \to T)\frac{\exp(-\beta E_n(0))}{%
Z_0(\beta; \{J_{ij}\})}  \nonumber \\
&=& \sum_{m,n} \frac{\mathrm{e}^{-\beta E_{m}(T) }}{Z_0(\beta; \{J_{ij}\})}%
P_{m,n}(0 \to T)  \nonumber \\
&=& \sum_{m} \frac{\mathrm{e}^{-\beta E_{m}(T) }}{Z_0(\beta; \{J_{ij}\})} 
\nonumber \\
&=& \frac{Z_T(\beta; \{J_{ij}\}) }{Z_0(\beta; \{J_{ij}\})},  \label{QAJE1}
\end{eqnarray}
where we used the fact that the performed work $W$ is a classical number and 
\begin{eqnarray}
\sum_{n}P_{m,n}(0 \to T) &=& \sum_{n} \langle \Psi_m(T)| U_T | \Psi_n(0)
\rangle \langle \Psi_n(0)| U^{\dagger}_T | \Psi_m(T) \rangle  \nonumber \\
&=& \sum_m \langle \Psi_m(T)| U_T U^{\dagger}_T | \Psi_m(T) \rangle = 1.
\end{eqnarray}
If we measure the physical observable $\hat{O}_T$ at the last of the
nonequilibrium process, we obtain another equation as 
\begin{equation}
\langle \hat{O}_{T}\mathrm{e}^{-\beta W}\rangle_{\mathrm{QA}} = \langle \hat{%
O} \rangle_{\beta} \frac{Z_T(\beta; \{J_{ij}\})}{Z_0(\beta; \{J_{ij}\})},
\label{QAJE2}
\end{equation}
where the subscript on the square brackets in the right-hand side denotes
the thermal average in the last equilibrium state with the inverse
temperature $\beta$.

Below we show several observations by application of JE for the isolated
quantum system to implementation of NQA.

\subsubsection{Performance of non-adiabatic quantum annealing}

First we discuss the possibility of NQA as a solver. Let us consider to
measure equilibrium quantities through NQA. The ratio of Eqs. (\ref{QAJE1})
and (\ref{QAJE2}) gives 
\begin{equation}
\frac{\langle \hat{O}_T\mathrm{e}^{-\beta W}\rangle_{\mathrm{QA}}}{\langle%
\mathrm{e}^{-\beta W}\rangle _{\mathrm{QA}}} =\langle \hat{O}
\rangle_{\beta}.  \label{QAJE3}
\end{equation}
The resultant equation suggests that the thermal average under the
Hamiltonian $H_0$ on the right-hand side can be estimated by NQA on the
left-hand side. This fact may be useful in the evaluation of equilibrium
average, since the left-hand side is evaluated without slow adiabatic
processes. In order to investigate the property of the ground state, we tune
the inverse temperature into a very large value $\beta \gg 1$. We should be
careful because the average on the left-hand side involves a non-extensive
quantity, the exponentiated work, whose value fluctuates significantly from
process to process. The average on the left-hand side must be calculated by
many trials of annealing processes. Thus, rare events with large values of
the exponentiated work (i.e. $\beta |W|\gg \Gamma _{0}$) would contribute to
the average significantly, and we have to repeat the annealing process very
many times in order to reach the correct value of the average. It usually
needs very many, typically exponentially many, repetitions. Thus the
difficulty has not been relaxed yet in general, but the present new
perspective may lead to different methods and tools than conventional ones
to attack the problem.

\subsection{Analyses on non-adiabatic quantum annealing}

Unfortunately, we have not reached any answers on the performance of NQA.
Instead We here evaluate several properties in nonequilibrium process as in
NQA for the particular spin glasses. We can exactly analyze nonequilibrium
behavior by combination of JE with the gauge transformation, although, in
general, there are few exact results in nonequilibrium quantum dynamical
system with many components.

Following the prescription of the Jarzynski equality, we consider a
repetition of NQA starting from the equilibrium ensemble. Let us remember
the whole Hamiltonian of QA for the typical spin glasses. The initial
Hamiltonian is given only by the transverse field $H(0) = H_1$, which means
a trivial uniform distribution. Consequently, as a starting point of our
analyses, we write down the specialized JE to the case for NQA as 
\begin{equation}
\langle \mathrm{e}^{-\beta W}\rangle _{\mathrm{QA}}=\frac{Z_{T
}(\beta,\{J_{ij}\})}{(2\cosh \beta \Gamma _{0})^{N}}.  \label{QAJE}
\end{equation}
We assume that the interactions $\{J_{ij}\}$ follow the distribution
function for the $\pm J$ Ising model (\ref{pmJ}), which is better to be
rewritten as 
\begin{equation}
P(J_{ij}) = \frac{\exp(\beta_p J_{ij})}{2\cosh \beta_p J},
\end{equation}
where we do not use $K=\beta J$ for transparency, and $\exp(-2\beta_p
J)=(1-p)/p$.

\subsubsection{Gauge transformation for quantum spin systems}

For several special spin glasses as the $\pm J$ Ising model, the gauge
transformation is available for analyses on the dynamical property even
under quantum fluctuations. The time-dependent Hamiltonian as in Eq. (\ref%
{QAS}) is invariant under the following local transformation, 
\begin{equation}
\sigma_i^x\to \sigma_i^x,~\sigma_i^y\to \xi_i\sigma_i^y,~\sigma_i^z\to
\xi_i\sigma_i^z,~J_{ij} \to J_{ij}\xi_i\xi_j\quad (\forall i,j),
\end{equation}
where $\xi_i (=\pm 1)$ is called as a gauge variable. This transformation is
designed to preserve the commutation relations between different components
of Pauli matrix \cite{Morita}. We skillfully use the gauge transformation to
analyze the dynamical property of the nonequilibrium behavior in NQA.

\subsubsection{Relationship between two different paths of NQA}

\begin{figure}[tb]
\begin{center}
\includegraphics[width=80mm]{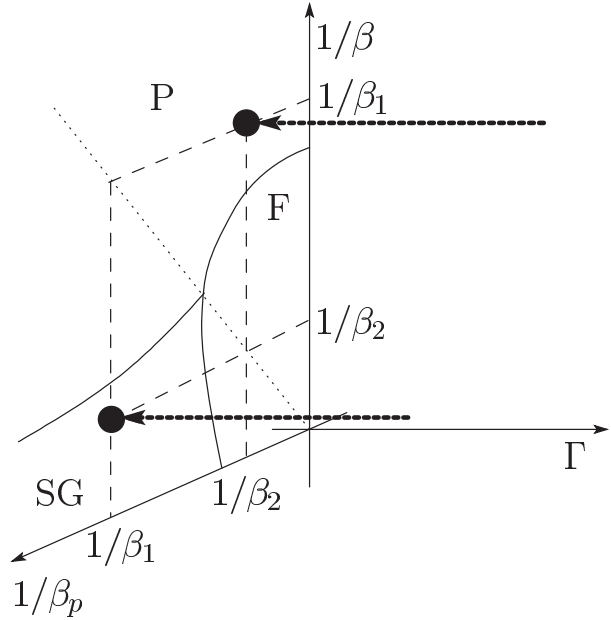}
\end{center}
\caption{{\protect\small Two different processes of NQA in Eq. (\protect\ref%
{QAdiff}). The left-hand side of Eq. (\protect\ref{QAdiff}) represents the
annealing process ending at the upper-right black dot and the right-hand
side terminates at the lower-left dot. Three phases expressed by the same
symbols as in Fig. \protect\ref{PG} are separated by solid curves and a
vertical line. The dotted line expresses the NL $\protect\beta_p = \protect%
\beta$. }}
\label{NQAfig1}
\end{figure}

Below, we reveal several properties inherent in NQA by the gauge
transformation. Let us take the configurational average of Eq. (\ref{QAJE})
over all the realizations of $\{J_{ij}\}$ for the special case with $\beta
=\beta_1$ and $\beta_p=\beta_2$ as 
\begin{equation}
\left[ \langle \mathrm{e}^{-\beta_1 W}\rangle _{\mathrm{QA}}\right] _{\beta
_2}=\left[ \frac{Z_{T }(\beta_1;\{J_{ij}\})}{\left( 2\cosh \beta_1 \Gamma
_{0}\right) ^{N}}\right] _{\beta _2}.
\end{equation}
The right-hand side is written explicitly as 
\begin{equation}
\left[ \langle \mathrm{e}^{-\beta_1 W}\rangle _{\mathrm{QA}}\right]
_{\beta_2}=\sum_{\{J_{ij}\}}\frac{\exp \big(\beta_2\sum_{\langle ij\rangle
}J_{ij}\big)}{(2\cosh \beta_2 J)^{N_{B}}}\frac{Z_{T}(\beta_1 ;\{J_{ij}\})}{%
\left( 2\cosh\beta_1 \Gamma _{0}\right) ^{N}}.
\end{equation}%
Let us here apply the gauge transformation and sum over all possible
configurations of the gauge variables $\{\xi _{i}\}$. We obtain, after
dividing the result by $2^{N}$, 
\begin{equation}
\left[ \langle \mathrm{e}^{-\beta_1 W}\rangle _{\mathrm{QA}}\right]
_{\beta_2}=\sum_{\{J_{ij}\}}\frac{Z_{T}(\beta_2
;\{J_{ij}\})Z_{T}(\beta_1;\{J_{ij}\})}{2^{N}(2\cosh \beta_2J)^{N_{B}}\left(
2\cosh \beta_1 \Gamma _{0}\right) ^{N}}.  \label{NQA1}
\end{equation}
A similar quantity of the average of the exponentiated work for the spin
glass with the inverse temperature $\beta _2$ and the parameter for the
quenched randomness $\beta_1 $ gives 
\begin{equation}
\left[ \langle \mathrm{e}^{-\beta_2W}\rangle _{\mathrm{QA}}\right] _{\beta_1
}=\sum_{\{J_{ij}\}}\frac{Z_{T}(\beta_2;\{J_{ij}\})Z_{T}(\beta_1;\{J_{ij}\})}{%
2^{N}(2\cosh \beta_1 J)^{N_{B}}\left( 2\cosh \beta_2 \Gamma _{0}\right) ^{N}}%
.  \label{NQA2}
\end{equation}
Comparing Eqs. (\ref{NQA1}) and (\ref{NQA2}), we find the following relation
between two different non-adiabatic processes, 
\begin{equation}
\left[ \langle \mathrm{e}^{-\beta_1 W}\rangle _{\mathrm{QA}}\right]
_{\beta_2}=\left[ \langle \mathrm{e}^{-\beta_2 W}\rangle _{\mathrm{QA}}%
\right] _{\beta_1}\left( \frac{\cosh \beta_1 J}{\cosh \beta_2 J}\right)
^{N_{B}}\left( \frac{\cosh \beta_2 \Gamma _{0}}{\cosh \beta_1 \Gamma _{0}}%
\right) ^{N}.  \label{QAdiff}
\end{equation}
We describe the two different paths of NQA related by this equality in Fig. %
\ref{NQAfig1}. 
Setting $\beta_2=0$ in Eq. (\ref{QAdiff}), (implying $p=1/2$, the symmetric
distribution or the high-temperature limit), we find a simple equality on
the performed work during NQA 
\begin{equation}
\left[ \langle \mathrm{e}^{-\beta_1 W}\rangle _{\mathrm{QA}}\right] _{0}=%
\frac{(\cosh \beta_1 J)^{N_{B}}}{(\cosh \beta_1 \Gamma _{0})^{N}}.
\label{QA01}
\end{equation}
The symmetric distribution ($\beta_2=0$ on the left-hand side) makes it
possible to reduce the right-hand side to the above trivial expression. It
is remarkable that NQA, which involves very complex dynamics, satisfies such
a simple identity irrespective of the speed of annealing $T$. If we apply
the Jensen inequality to the above equality, we can obtain the lower bound
for the performed work as 
\begin{equation}
\left[ \langle W\rangle _{\mathrm{QA}}\right]_{0} \ge - \frac{N}{\beta}
\log\left( \frac{(\cosh \beta J)^{\frac{N_B}{N}}}{\cosh \beta \Gamma _{0}}%
\right).  \label{QAWorkL}
\end{equation}
Here we generalize the inverse temperature to $\beta$ from the specific
choice $\beta_1$. This lower bound is loose, since the direct application of
the Jensen inequality to JE for NQA yields, after the configurational
average with the symmetric distribution, 
\begin{equation}
\left[ \langle W\rangle _{\mathrm{QA}}\right]_{0} \ge \frac{1}{\beta}%
D(0|\beta)- \frac{N}{\beta} \log\left( \frac{(\cosh \beta J)^{\frac{N_B}{N}}%
}{\cosh \beta \Gamma _{0}}\right),  \label{QAWorkB}
\end{equation}
where $D(\beta|\beta^{\prime })$ is the Kullback-Leibler divergence defined
as 
\begin{equation}
D(\beta|\beta^{\prime }) = \sum_{\{J_{ij}\}} \tilde{P}_{\beta}(\{J_{ij}\})
\log \frac{\tilde{P}_{\beta^{\prime }}(\{J_{ij}\})}{\tilde{P}%
_{\beta}(\{J_{ij}\})}.
\end{equation}
Here we defined the marginal distribution for the specific configuration $%
\{J_{ij}\}$ summed over all the possible gauge transformations, 
\begin{equation}
\tilde{P}_{\beta}(\{J_{ij}\}) =\frac{1}{2^N} \sum_{\{\xi_i\}}\prod_{\langle
ij\rangle} P(J_{ij}) = \frac{Z_{T}(\beta;\{J_{ij}\})}{2^N(2\cosh \beta
J)^{N_B}}.
\end{equation}
Since the Kullback-Leibler divergence does not become non-negative, the work
performed by the transverse field during a nonequilibrium process in in the
symmetric distribution (i.e. the left-hand side of Eq. (\ref{QAWorkB})) does
not exceed the second quantity on the right-hand side of Eq. (\ref{QAWorkB}%
). This fact means Eq. (\ref{QAWorkL}) was looser.

\subsubsection{Exact relations involving inverse statistics}

\begin{figure}[tb]
\begin{center}
\includegraphics[width=80mm]{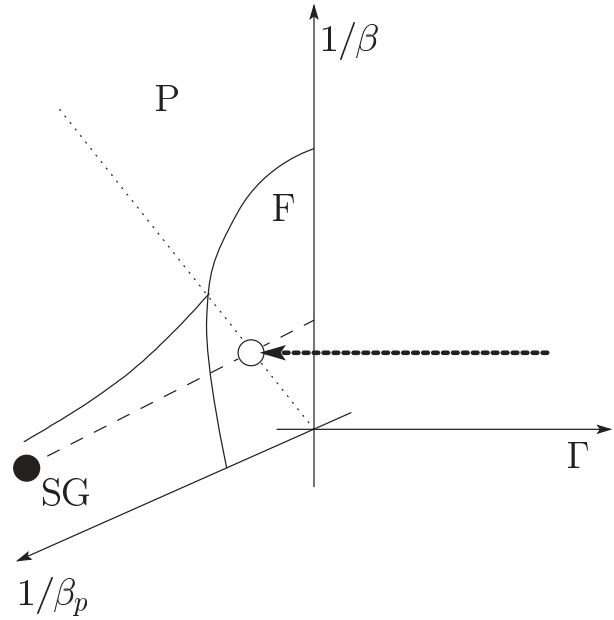}
\end{center}
\caption{{\protect\small Two different nonequilibrium processes in NQA
through Eq. (\protect\ref{IQAdiff}). We use the same symbols as in Fig. 
\protect\ref{PG}. The white circle denotes the target of the process on the
right-hand side of Eq. (\protect\ref{IQAdiff}), whereas the black dot is for
the left-hand side. }}
\label{NQAfig2}
\end{figure}

Beyond the above results, we can perform further non-trivial analyses for
the nonequilibrium process in the special conditions. Let us next take the
configurational average of the inverse of the Jarzynski equality, Eq. (\ref%
{QAJE}), as 
\begin{equation}
\left[\frac{1}{\langle \mathrm{e}^{-\beta W} \rangle_{\mathrm{QA}}}\right]%
_{\beta_p} = \left[\frac{\left( 2 \cosh \beta\Gamma_0 \right)^{N}}{%
Z_{T}(\beta;\{J_{ij}\})}\right]_{\beta_p}.  \label{QAinverse0}
\end{equation}
The application of the gauge transformation to the right-hand side yields 
\begin{equation}
\left[\frac{1}{\langle \mathrm{e}^{-\beta W} \rangle_{\mathrm{QA}}}\right]
_{\beta_p} = \sum_{\{J_{ij}\}}\frac{\exp\big(\beta_p\sum_{\langle ij
\rangle}J_{ij}\xi_i\xi_j\big)}{(2\cosh \beta_p J)^{N_B}}\frac{\left( 2 \cosh
\beta\Gamma_0 \right)^{N}}{Z_{T}(\beta;\{J_{ij}\})}.
\end{equation}
By summing the right-hand side over all the possible configurations of $%
\{\xi_i\}$ and dividing the result by $2^N$, we reach 
\begin{equation}
\left[\frac{1}{\langle \mathrm{e}^{-\beta W} \rangle_{\mathrm{QA}}}\right]%
_{\beta_p} = \sum_{\{J_{ij}\}}\frac{Z_{T}(\beta_p;\{J_{ij}\})}{2^N(2\cosh
\beta_p J)^{N_B}}\frac{\left( 2 \cosh \beta\Gamma_0 \right)^{N}}{%
Z_{T}(\beta;\{J_{ij}\})}.  \label{IQA}
\end{equation}
If we set $\beta_p = \beta$ on the Nishimori line, this equation reduces to 
\begin{equation}
\left[\frac{1}{\langle \mathrm{e}^{-\beta W} \rangle_{\mathrm{QA}}}\right]%
_{\beta} = \frac{(\cosh \beta\Gamma_0)^N}{(\cosh \beta J)^{N_B}}.
\label{QA02}
\end{equation}
Comparison of Eqs. (\ref{QA01}) and (\ref{QA02}) leads us to 
\begin{equation}
\left[\langle \mathrm{e}^{-\beta W} \rangle_{\mathrm{QA}}\right]_{0} = \left(%
\left[\frac{1}{\langle \mathrm{e}^{-\beta W} \rangle_{\mathrm{QA}}}\right]
_{\beta}\right)^{-1}.  \label{IQAdiff}
\end{equation}
As shown in Fig. \ref{NQAfig2}, two completely different processes are
nontrivially related by the resultant relation: One toward the Nishimori
line and the other for the symmetric distribution. 

Let us further consider the inverse of Eq. (\ref{QAJE2}) for the two-point
correlation $O_T = \sigma^z_i\sigma^z_j$. We take the configurational
average of both sides under the condition $\beta_p = \beta $ as 
\begin{equation}
\left[\frac{1}{\langle \sigma^z_i\sigma^z_j\mathrm{e}^{-\beta W} \rangle_{%
\mathrm{QA}}}\right]_{\beta} = \frac{(\cosh \beta\Gamma_0)^N}{(\cosh \beta
J)^{N_B}}\left[\frac{1}{\langle\sigma^z_i\sigma^z_j\rangle_{\beta}}\right]%
_{\beta}.
\end{equation}
The quantity on the right-hand side becomes unity by the gauge
transformation as has been shown in the literatures \cite{HN81,HNbook}. We
thus obtain a simple exact relation 
\begin{equation}
\left[\frac{1}{\langle \sigma^z_i\sigma^z_j\mathrm{e}^{-\beta W} \rangle_{%
\mathrm{QA}}}\right]_{\beta} = \frac{(\cosh \beta\Gamma_0)^N}{(\cosh \beta
J)^{N_B}},
\end{equation}
which is another exact identity for processes of NQA.

The importance of the above equalities is still not clear. However it is
true that we have little exact results on the nonequilibrium behavior of the
spin glasses driven by quantum fluctuations as the transverse field. When we
realize the quantum spin systems in experiments, the above results can play
a roll as their indicators to verify their precisions and conditions. We
should emphasize that the specialized tool to analyze spin glasses has
facilitated the dynamical property in nonequilibrium process driven by
quantum fluctuations. Beyond our analyses introduced here, we hope that more
fascinating results on quantum computations would be obtained in the future
with recourse to several techniques developed in statistical mechanics.

\section{Summary}

\label{aba:sec5} We looked over two topics lying between quantum information
processing and statistical mechanics.

The first was the quantum error correction using the property of the
topology. We prepare the redundant physical qubits in order to express the
logical qubits we encode the specific information. Although we should deal
with the quantum-many body systems, the used technique was proposed from the
classical method in statistical mechanics. Statistical mechanics is
available to facilitate to identify its precise location of the theoretical
limitation to successfully infer the original state. In this direction with
rich knowledge in statistical mechanics, we will be able to propose another
ingenious way to give the quantum state more robustness and resilience as in
the case of the depolarizing channel.

The second part was to improve the performance of quantum annealing away
from the adiabatic-control case. The theoretical key was the Jarzynski
equality. We suggest two ways to overcome the particular bottleneck in the
adiabatic computation of quantum annealing. Both of the schemes would be
needed for many-time repetition to produce the desired results. However we
hope that the studies in this direction would give a novel technique beyond
the ordinary limitations. The key point will be to enhance the possibility
with the desired states. In the classical counterpart, recently several
techniques are proposed with use of the population to reach the desired
distribution, that is equilibrium state \cite{NJ,Pop1,Iba,ON}. Such a
technique as in the classical case enables us to generate the particular
quantum state with higher probability than expected by use of skillful
techniques.

The bridge between the quantum information processing and statistical
mechanics continues to lead us to frontiers, where we encounter novel and
surprising results beyond expected ones from knowledge we obtain
only from each side. We are at a position to look at the birth of such a
fascinating interdisciplinary field. Don't miss it.

\bibliographystyle{ws-procs9x6}
\bibliography{ws-pro-sample}

\end{document}